\documentclass[10pt]{article}
\usepackage{geometry}
\usepackage{textgreek}
\usepackage{amsfonts}
\usepackage{amsmath}
\usepackage{mathtools}
\usepackage{slashed}
\usepackage{amssymb}
\usepackage{cancel}
\usepackage{mathrsfs}
\usepackage{wasysym}
\usepackage{physics}
\usepackage{cite}
\usepackage{bbm}
\usepackage{bm}
\usepackage{graphicx}
\usepackage[labelfont=small]{caption}
\usepackage{hyperref}
\usepackage[usenames,dvipsnames]{color}
\usepackage{wrapfig}
\usepackage{comment}
\usepackage{multicol}
\usepackage{cleveref}
\usepackage[normalem]{ulem}

\newcommand{\sfrac}[2]{ {\textstyle \frac{#1}{#2}}}
\newcommand{\st}[1]{\text{\tiny \rm #1}}

\newcommand{\1}{\hat{1}}

\newcommand{\x}{\hat{x}}
\newcommand{\A}{\hat{a}}
\newcommand{\La}{\hat{\Lambda}}
\newcommand{\e}{\hat{e}}
\newcommand{\E}{\hat{E}}
\newcommand{\y}{\hat{y}}

\newcommand{\p}{\xi}

\newcommand{\mP}{\hat P}
\newcommand{\mM}{\hat M}

\newcommand{\X}{\hat{\xi}}
\newcommand{\Ch}{\hat{\chi}}
\newcommand{\M}{\hat{M}}

\hypersetup{
colorlinks=true,         % false: boxed links; true: colored links, false is default
citecolor=blue,           % color of links to bibliography
}

\title{\vspace{-1.5cm} \Large \bf 
Covariant quantization of field theories on\\
T-Minkowski noncommutative spacetimes}
\date{}
\author{{$^{(1,2,3)}$Giuseppe Fabiano\footnote{gfabiano@lbl.gov}, $^{(4)}$Flavio Mercati\footnote{flavio.mercati@gmail.com}}
\vspace{12pt}
\\
\small $^{(1)}$Physics Division, Lawrence Berkeley National Laboratory, Berkeley, CA 94720, USA\\
\small $^{(2)}$Department of Physics, University of California, Berkeley, CA 94720, USA\\
\small $^{(3)}$Centro Ricerche Enrico Fermi, I-00184 Rome, Italy\\
\small $^{(4)}$Departamento de F\'isica, Universidad de Burgos, 09001 Burgos, Spain.
}

\begin{document}
\maketitle
\begin{abstract}
    We develop a quantization scheme for the quantum theory of a real scalar field on a class of non-commutative spacetime models collectively known as T-Minkowski. Requiring the theory to be covariant under T-Poincaré transformations, we find that for a subclass of models the Wightmann functions are equal to their commutative counterparts, and we are able to prove a Wick theorem for Wightmann functions that is structurally equivalent to the one encountered in commutative QFT. For some of these models we further extend the result to Green functions and to $N$-point functions of interacting QFT, which we also find to be commutative, leaving no space for IR/UV mixing effects advocated in other approaches to noncommutative QFT.
\end{abstract}
\tableofcontents

\newpage

\subsection*{Notations used in the paper}
\begin{itemize}
\item The Einstein summation convention is assumed. Greek indices $\alpha, \beta , \dots \mu,\nu,\dots$ will run from $0$ to $3$. Lowercase Latin from the beginning of the alphabet, $a,b,c,\dots$ will be natural numbers.

\item Symmetrization and antisymmetrization of indices will be done with a $1/2$ weight:
$$
T^{(\mu\nu)} = {\sfrac 1 2} (T^{\mu\nu} + T^{\nu\mu} ) \,,
\qquad
T^{[\mu\nu]} = {\sfrac 1 2} (T^{\mu\nu} - T^{\nu\mu} ) \,.
$$

\item A hat on a symbol, $\hat f$, indicates that $\hat f$ is an element of a not-necessarily-commutative algebra.

\item $\eta_{\mu\nu}$ will indicate a \textit{numerical,} constant symmetric metric. The Minkowski metric in Cartesian coordinates, \textit{i.e.} $\eta_{\mu\nu} = \text{diag}(-1,1,1,1)$ will be assumed unless specified otherwise. The signature is the mostly-positive one.

\item Complex conjugates will be indicated with a bar, $\overline{x+iy} = x-iy$, $x,y \in \mathbbm{R}$. The involution of our noncommutative algebras (an antilinear antihomomorphism) will be represented by a $(\,\cdot \,)^*$ symbol, to distinguish it from the Hermitian conjugate symbol $(\,\cdot \,)^\dagger$, which will be reserved for the creation and annihilation operators of noncommutative QFT.
\end{itemize}

\section{Introduction}\label{intro}

Quantum Field Theory (QFT) on noncommutative spacetimes has been studied for at least three decades, mainly motivated by quantum gravity arguments, which indicate that the locality principle should be relaxed or violated at small distances approaching the Planck scale~\cite{Vitale:2023znb}. Assuming that  spacetime acquires the properties of a noncommutative geometry~\cite{Connes_book94} at short distances is viewed as a promising way to have an effective, large-scale description of the Planck-scale ``spacetime foam'' structure suggested by quantum gravity arguments~\cite{MisnerWheeler}. Such a microscopic structure of empty space is expected to embed a length scale (presumably closely related to, if not identical to, the Planck length) into the very fabric of spacetime, and this in turn raises the issue of the fate of Lorentz invariance. Violations of the Special Relativity Principle (\textit{i.e.} the equivalence between inertial reference frames) are constrained with extreme levels of precision~\cite{Mattingly:2005re,Bolmont:2022yad}, so it would be very hard, for a Lorentz-breaking theory, to avoid all different experimental constraints on Lorentz violation, even if the effect is controlled by the tiny Planck length.

An intriguing alternative possibility is that a noncommutative QFT remains compatible with the Special Relativity Principle, rendering powerless any constraint based on effects that are sensitive to the state of motion with respect to some ether's preferred frame (\textit{i.e.} Michelson--Morley-type experiments). Such a model needs to reconcile the existence of a fundamental Length scale with the equivalence between reference frames, and it therefore needs to implement Lorentz/Poincaré invariance in a non-standard way~\cite{AmelinoCamelia:2000mn}.
The theory of quantum groups (Hopf algebras)~~\cite{majid_1995,majid_2002,Chari} provides precisely an example of such construction, where a Lie group is promoted to a noncommutative geometry (\textit{i.e.} the group manifold is noncommutative), and its homogeneous spaces are noncommutative spaces with a deformed notion of isometry. A QFT built on one such construction, in which Minkowski space is noncommutative and is covariant under a quantum-group deformation of the Poincaré group, would fit all the requirements for a promising effective theory of matter on a non-classical background, in which the quantum gravity effects are averaged into a large-scale noncommutative description. There have been many attempts at such a construction, but so far all come short of displaying a fully consistent QFT whose physical predictions are demonstrably Poincaré covariant.

In recent investigations \cite{Mercati:2023apu,Mercati:2024rzg}, one of us introduced the so-called T-Minkowski spacetimes, a family of noncommutative geometries whose isometries are described by a quantum-group deformations of the Poincaré algebra, the corresponding Hopf algebras being collectively called T-Poincaré.  In these works, all the essential mathematical tools for developing  field theories on these spacetimes were developed, including noncommutative differential calculus, braided N-point algebras, and Fourier theory. This enables rigorous definitions and analyses of classical fields.
The classical theories thereby obtained are T-Poincaré invariant, in the sense of invariant under an action of the deformed T-Poncaré groups that describe the isometries of T-Minkowski spacetimes.
The question remains whether these models allow for the  consistent definition of a T-Poincaré-covariant QFT. This step is non trivial, as a straightforward quantization of the classical models can easily lead to a Lorentz-violating model. For example, two of the simplest T-Minkowski models are the so-called theta and rho models ($\theta$-Minkowski and $\varrho$-Minkowski), defined, respectively, by the commutation relations:
\begin{equation}
[\x^\mu,\x^\nu]=i \theta^{\mu\nu} \,,
\end{equation}
where $\theta^{\mu\nu}$ is an antisymmetric numerical matrix (supposed multiplied by the identity of the algebra, in the commutators), and,  for $\rho$-Minkowski:
\begin{equation}
[\x^0,\x^1]=i \x^2 \,, \quad [\x^0,\x^2]=-i \x^1  \,,
\qquad [\x^1,\x^2]= [\x^0,\x^3]=[\x^1,\x^3]=[\x^2,\x^3]=0\,.
\end{equation}
With a natural path-integral quantization, both models above present a feature called IR/UV mixing. Namely, in certain processes (\textit{e.g.} the four-point function of the free theory, or the one-loop corrections to the propagator of a $\lambda \, \varphi^4$ theory), some of the contributions to the amplitudes have Lorentz-violating factors that do not conserve the total momentum (signaling also a breakdown of translation invariance). This has been shown in $\theta$-Minkowski in~\cite{Minwalla:1999px}, and in $\rho$-Minkowski in~\cite{Hersent:2023lqm}. The same phenomenon is present also in the much-studied $\kappa$-Minkowski spacetime~\cite{Amelino-Camelia:2001rtw,Hersent:2024beg}, although only its ``lightlike'' version qualifies as belonging to the T-Minkowski class of models. 
The IR/UV mixing phenomenon is so called because the Feynman diagrams that suffer from it have momentum-dependent phases whose rapid oscillations regularize the UV divergences of the momentum integrals, at the expenses of the infrared behaviour, where new divergences appear in the zero momentum limit.  This behaviour has generally been considered pathological, and evading it was the central motivation of important results like~\cite{Grosse:2003nw}, which provides one of the few examples of interacting QFT that is well-defined at all energies (and, incidentally, is a Poincaré-violating noncommutative QFT on the 2D version of $\theta$-Minkowski spacetime). All these results, including~\cite{Grosse:2003nw}  which explicitly breaks translation invariance through a background potential, ignore the issue of the relativistic invariance of the theory, which for us is the most problematic aspect of IR/UV mixing. The present paper is focused on the (deformed) Poincaré invariance of QFTs on T-Minkowski noncommutative spacetimes, and intends to answer the question of whether there exists a quantization scheme that preserves relativistic invariance. The answer is positive, at least for a class of simpler models. In particular, we can prove the existence of a covariant quantization scheme for all \textit{unimodular} models of~\cite{Mercati:2023apu} in the case of a non-interacting scalar field. These are models 10--16 and 19--21 of~\cite{Mercati:2023apu}. In the case of general interacting scalar fields, we can give a positive (perturbative) answer only for models 13--15 and 19--21 (so-called $\zeta$- and $\theta$-Minkowski). Notice that, in the case of $\theta$-Minkowski (models 19--21) our results coincide with those obtained in~\cite{oeckl2000untwisting,Fiore:2007vg}. Indeed our quantization scheme follows the one introduced in~\cite{Fiore:2007vg}, and generalizes it to all unimodular T-Minkowski models.

\section{Mathematical preliminaries on T-Minkowski models}
\label{matprel}
The main results of the paper in \Cref{sec:ncqft} will be derived for unimodular T-Minkowski models. However, the mathematical tools developed in \cite{Mercati:2023apu,Mercati:2024rzg} and some new technical results we will show in this section are valid for any T-Minkowski model, unless otherwise specified.

\subsection{T-Poincaré, T-Minkowski and the braided tensor product algebra}
\label{sec:TMin_introduction}

Consider the Poincaré group  $ISO(3,1)$, and a noncommutative deformation of the  algebra of functions on the group, which we will call $C_\ell[ISO(3,1)]$. The generators of this algebra are coordinate functions, denoted by $\hat{\Lambda}^{\mu}{}_\nu$ and $\hat{a}_\mu$ 
with the $\La^\mu{}_\nu$ satisfying the constraint of being Lorentz matrices:
\begin{equation}
\label{MetricInvariance}
    \eta_{\mu\nu}  \,   \La^\mu{}_\rho \, \La^\nu{}_\sigma 
= \eta_{\rho\sigma} 
\,, \qquad
\eta^{\rho\sigma}  \,   \La^\mu{}_\rho \, \La^\nu{}_\sigma 
= \eta^{\mu\nu} \,.
\end{equation}
The matrix $\eta^{\mu\nu}$ is any real nondegenerate symmetric matrix with signature $(-,+,+,+)$. For the present paper, we will assume the form $\eta_{\mu\nu} = \text{diag}(-1,1,1,1)$ for the metric.
The group product law is written, in this language, as a coproduct map:
\begin{equation}
    \Delta (\hat{\Lambda}^\mu{}_\nu) = \hat{\Lambda}^\mu{}_\rho \otimes \hat{\Lambda}^\rho{}_\nu \,, ~~ \Delta (\hat{a}^\mu) = \hat{\Lambda}^\mu{}_\nu \otimes \hat{a}^\nu \,,
\end{equation}
and an antipode and counit maps encode the group inverse and the identity element, respectively:
\begin{equation}
    S(\hat{\Lambda}^\mu{}_\nu) = (\hat{\Lambda}^{-1})^\mu{}_\nu \,, \qquad S(\hat{a}^\mu) = -\hat{a}^\nu(\hat{\Lambda}^{-1})^\mu{}_\nu \,, \qquad  \epsilon (\hat{\Lambda}^\mu{}_\nu) = \delta^\mu{}_\nu \,,\qquad \epsilon(\hat{a}^\mu)=0
\end{equation}
So far we only have a description of the Lie group structures. In order to define a quantum group, we need to deform the pointwise product between functions on the group into a noncommutative product, which needs to be compatible with all the other structures so far defined on $C_\ell[ISO(3,1)]$~\cite{wess1999qdeformed}. In~\cite{Mercati:2023apu}, a set of physical and simplicity assumptions were used to narrow down the ansatz for this product to
\begin{equation}\label{T-PoincareGroup_commutators}
\left\{\begin{aligned}
  &[ \A^\mu , \A^\nu ]  =  
 i \, \left( f^{\nu\mu}{}_\rho - f^{\mu\nu}{}_\rho \right) \, \A^\rho  \,,
\\
&[ \La^{\mu}{}_\rho  , \A^{\nu} ]   =
  i \,  f^{\alpha\beta}{}_\gamma
\left(  \La^{\mu}{}_\alpha \, \La^{\nu}{}_\beta \, \delta^\gamma{}_\rho  - \delta^{\mu}{}_\alpha \, \delta^{\nu}{}_\beta \, \La^\gamma{}_\rho \right) 
 \,, 
\\
&[ \La^{\mu}{}_\rho ,  \La^{\nu}{}_\sigma ] = 0   \,,
 \end{aligned}\right.
\end{equation}
where the coefficients $ f^{\alpha\beta}{}_\gamma$ are real ``structure constants'', satisfying $f^{\alpha\beta\gamma}=-f^{\gamma\beta\alpha}$, and $\theta^{\alpha\beta}=-\theta^{\beta\alpha}$ are real as well. Together, they satisfy the following constraints
\begin{equation}  \label{StructureConstants_constraints}
     \begin{gathered}
      f^{[\mu\nu]}{}_\lambda \, f^{[\lambda\rho]}{}_\sigma - 
        \, f^{[\mu\rho]}{}_\lambda f^{[\lambda\nu]}{}_\sigma 
        - f^{[\nu\rho]}{}_\lambda \, f^{[\mu\lambda]}{}_\sigma  = 0\,,
        \qquad 
 f^{[\mu\nu]}{}_\lambda \, \theta^{\rho \lambda}   + f^{[\rho\mu]}{}_\lambda \, \theta^{\nu \lambda} + f^{[\nu\rho]}{}_\lambda \, \theta^{\mu \lambda} =  0\,,
\\
    f^{\alpha\mu}{}_\lambda  \, f^{\lambda\nu}{}_\beta -      f^{\alpha\nu}{}_\lambda  \, f^{\lambda\mu}{}_\beta 
    = 2f^{[\mu\nu]}{}_\rho \, f^{\alpha\rho}{}_\beta \,,
    \qquad  
    \eta^{\theta (\sigma}  \,  f^{\varphi ) \lambda }{}_\theta =  0 \,.
     \end{gathered}
\end{equation}
The solutions to the above equation (in the case of the Poincar\'e algebra) have been classified by Zakrzewski~\cite{Zakrzewski_1997} and Tolstoy in \cite{tolstoy2007twistedquantumdeformationslorentz}. The classification leads to 16 classes (Cases \textbf{7}, \textbf{8}, \textbf{9}$^{\bm{(-1)}}$, \textbf{9}$^{\bm{(0)}}$,\textbf{9}$^{\bm{(+1)}}$, \textbf{10}--\textbf{21} in~\cite{Zakrzewski_1997}) of inequivalent models (up to automporphisms of the Poincaré group) which are listed in \cref{app:Rmatrices} in terms of the formalism presented above in \cite{Mercati:2023apu,Mercati:2024rzg}. Each solution identifies a noncommutative homogeneous space, whose algebra of functions over the space we indicate as $C_\ell[\mathbbm{R}^{3,1}]$ and whose noncommutative product is given by
\begin{equation}
 \label{x_comm_rel}
  [\x^{\mu} , \x^{\nu}] 
    = i \, \left( f^{\nu\mu}{}_\rho - f^{\mu\nu}{}_\rho \right) \, \x^\rho    \,.
\end{equation}
These homogeneous spaces are invariant under the corresponding quantum Poincaré  group, in the sense that a coaction map $( \, \cdot \, )' : C_\ell[\mathbbm{R}^{3,1}] \to  C_\ell[ISO(3,1)] \otimes C_\ell[\mathbbm{R}^{3,1}] $ is an algebra homomorphism for $C_\ell[\mathbbm{R}^{3,1}]$. In terms of the four independent Cartesian coordinates, the coaction reproduces the usual Poincar\'e transformation of Minkowski coordinates:
\begin{equation}\label{Coaction_spacetime_coordinates}
 \x'^\mu = \La^\mu{}_\nu \otimes \x^\nu + \A^\mu \otimes \1 \,.
\end{equation}
This coaction leaves the relations~\eqref{x_comm_rel} invariant (homomorphism property), in the sense that
\begin{equation}
 \label{x_comm_rel_primed}
  [\x'^{\mu} , \x'^{\nu}] 
    = i \, \left( f^{\nu\mu}{}_\rho - f^{\mu\nu}{}_\rho \right) \, \x'^\rho    \,,
\end{equation}
The properties of the T-Minkowski models \cite{Mercati:2023apu} allow for the existence of a 4-dimensional bicovariant differential calculus in the sense of Woronowicz~\cite{Woronowicz:1989}, defined by the commutation relations 
\begin{equation}
\label{xdx_comm_rel}
 [ \x^{\mu} , d\x^{\nu} ]  = i \,f^{\nu\mu}{}_\rho \, d \x^\rho  \,.
\end{equation}
which is covariant under the simultaneous coaction~\eqref{Coaction_spacetime_coordinates} and the transformation law of the differential $d \x'^\mu = \La^\mu{}_\nu \otimes d \x^\nu $, which in turn expresses the fact that the differential is translation-invariant but not Lorentz-invariant.
Finally, the T-Minkowski models also allow one to introduce a higher-dimensional covariant comodule over $C_\ell[ISO(3,1)]$, referred to as braided tensor product algebra and denoted by $C_\ell[\mathbbm{R}^{3,1}]^{\otimes_\ell N}$, with defining commutation relations given by
\begin{equation}
  \label{xaxb_comm_rel}
  [\x_a^\mu , \x_b^\nu] 
    =  
     -   i    \,   f^{(\mu\nu)}{}_\rho \,  \left( \x_a^\rho - \x_b^\rho   \right)   
 - i   \, f^{[\mu\nu]}{}_\rho\left(     \x_a^\rho  +    \x_b^\rho     \right)    \,, \qquad a,b=1,...,N\, .
\end{equation}
which are covariant under the coaction \eqref{Coaction_spacetime_coordinates}, which takes the same form for each copy in $C_\ell[\mathbbm{R}^{3,1}]^{\otimes_\ell N}$. 
This braided tensor product algebra $C_\ell[\mathbbm{R}^{3,1}]^{\otimes_\ell N}$ consists of the noncommutative deformation of the algebra of functions of $N$ different spacetime points. This is a crucial ingredient of Quantum Field Theory, the physical content of the theory being encoded in the form of the $N$-point functions.

Starting from \eqref{xaxb_comm_rel} one can also show that
\begin{equation}
\label{eq:commcoorddiff}
 [ \x^{\mu}_a , \Delta \x^{\nu}_{bc} ]  = i \,f^{\nu\mu}{}_\rho \, \Delta\x^{\rho}_{bc} \,, \qquad \Delta \x^\mu_{bc} = \x^{\mu}_b - \x^\mu_c\,,
\end{equation}
namely, the commutation relations between a single coordinate and a generic coordinate difference have the same form of the commutation relations between a single coordinate and the differential $d\x^\mu$, evocative of the interpretation that, in a loose sense, $d\x^\mu$ can be regarded as an ``infinitesimal'' difference between coordinates.
Moreover, as done in \cite{Lizzi:2021rlb,DiLuca:2022idu} for the case of lightlike $\kappa$-Minkowski, the commutation relations \eqref{xaxb_comm_rel}, can be conveniently rewritten in terms of center of mass and relative position coordinates, respectively defined as
\begin{equation}
\label{eq:comdiff}
    \x_{cm}^\mu\coloneqq \frac{1}{N}\sum_a \x_a^\mu \,, \qquad \y_a^\mu\coloneqq \x_a^\mu-\x_{cm}^\mu \,, \qquad \sum_{a=1}^N \y_a^\mu = 0 \,. 
\end{equation}
The two types of variables, $\x_{cm}^\mu$ and $\y_a^\mu$ close two subalgebras of the commutation relations \eqref{xaxb_comm_rel}:
\begin{equation}
\label{eq:comcommrels}
    \left\{\begin{aligned}&[\x_{cm}^\mu,\x_{cm}^\nu]=i(f^{\nu\mu}{}_\rho-f^{\mu\nu}{}_\rho)\x_{cm}^\rho \\
    &[\x_{cm}^\mu,\y_a^\nu]=if^{\nu\mu}{}_\rho \y_a^\rho \\
    &[\y_a^\mu,\y_b^\nu]=0 \, . 
    \end{aligned}\right.
    \qquad
    \begin{aligned}
    &\y_a^\mu \in   \mathcal{M}^{N-1} \sim \mathbbm{R}^{4(N-1)} \,,
    \\
    &\x_{cm}^\mu \in \mathcal{M}^{cm}_\ell \sim C_\ell[\mathbbm{R}^{3,1}] \,,
    \end{aligned}
\end{equation}
We observe that the above commutators have the structure of a semidirect product Lie algebra:
\begin{equation}
   C_\ell[\mathbbm{R}^{3,1}]^{\otimes_\ell N} =  \mathcal{M}^{cm}_\ell \ltimes \mathcal{M}^{N-1} \,,
\end{equation}
where the Abelian subalgebra of relative coordinates $\mathcal{M}^{N-1}$ is acted upon the non-Abelian subalgebra of center of mass coordinates $ \mathcal{M}^{cm}_\ell$.  Moreover, the latter subalgebra is isomorphic to the single-point algebra $C_\ell[\mathbbm{R}^{3,1}]$~\eqref{x_comm_rel}.

Importantly, coordinate differences commute with each other (as first observed in~\cite{Fiore:2007vg}, in the case of $\theta$-Minkowski/Moyal spacetimes):
\begin{equation}
    [\Delta \x_{ab},\Delta \x_{cd}]=0\, , \qquad  \forall \, a,b,c,d  \,.
\end{equation}
This will enable a crucial simplification in the physical interpretation of the noncommutative QFT developed in \cref{sec:ncqft}. 
To conclude this section, let us specify the action of the $*$ involution on the relevant structures of the T-Minkowski framework. When applied on noncommutative products, it has to invert the order of all the elements appearing in the noncommutative products. On the generators of the $C_\ell[\mathbbm{R}^{3,1}]$, $C_\ell[ISO(3,1)]$, $C_\ell[\mathbbm{R}^{3,1}]^{\otimes_\ell N}$ algebras and on the differential $d\x^\mu$, we define it as
\begin{equation}\label{Involutions}
(\x^\mu)^* = \x^\mu \,, \qquad (\A^\mu)^* = \A^\mu \,, \qquad (\La^\mu{}_\nu)^* = \La^\mu{}_\nu \,, \qquad (\x^\mu_a)^* = \x^\mu_a \,,  \qquad (d \x^\mu)^* =  d \x^\mu \,.
\end{equation}
In this way, all of the commutators involving the generators above remain invariant under the $*$ involution.

\subsection{Fourier transform and integral calculus}

The algebra $C_\ell[\mathbbm{R}^{3,1}]$ is generated by the identity operator $\1$ and the four coordinates $\x^\mu$, and, since the Poincar\'e--Birkhoff--Witt  property holds \cite{wess1999qdeformed}, a basis for it is given by all possible monomials with a certain ordering prescription. The scalar fields will be expanded in terms of noncommutative exponentials in the spacetime coordinates:\footnote{Throughout the paper, a noncommutative function, element of the algebra $C_\ell[\mathbbm{R}^{3,1}]$  (or the braided tensor product algebra  $C_\ell[\mathbbm{R}^{3,1}]^{\otimes_\ell N}$), will be indicated either as a map applied on the noncommutative coordinates, \textit{e.g.} $f(\x)$, or as a map with the hat, in case the dependence on the noncommutative coordinates is not explicit, \textit{i.e.} $\hat{f}$.}
\begin{equation}\label{GeneralOrderedExponentials}
\E[q] = ~ : e^{i \, q_\mu \, \x^\mu} : ~ = \sum_{n=0}^\infty  \frac{i^n}{n!} \, :(q_\mu \, \x^\mu)^n  :  \,, 
\end{equation}
where $: \, . \, :$ is a given ordering prescription, and the exponential is understood as a formal power series.

The product of two exponentials can be written as
\begin{equation}\label{ProductBetweenExponentials}
\E[p] \, \E[q] =\E[\Delta(p,q)] \,,
\end{equation} 
where $\Delta : \mathbbm{R}^4 \otimes \mathbbm{R}^4 \to \mathbbm{R}^4$ is the momentum composition law (which turns out to be the coproduct of the translation sector of the T-Poincaré \textit{algebra} $U_\ell[iso(3,1)]$ dual to the quantum group, when written in a particular basis~\cite{Mercati:2024rzg}. The coproduct is an associative map
\begin{equation}
\Delta[p,\Delta(q,k)] = \Delta[\Delta(p,q), k] \,.
\end{equation}
In the following, when $N$ arguments $p_1,\cdots,p_N$ are involved in the coproduct, we will write it as $\Delta[p_1,\cdots,p_N]$, exploiting the fact that the map $\Delta$ is associative.

The coordinates of the identity $\1$ are defined by the counit $\epsilon(p) = o \in \mathbbm{R}^4  $ which is simply the null vector $o_\mu=(0,0,0,0)$. The counit is neutral with respect to the coproduct, meaning that
\begin{equation}
\Delta(o,p) = \Delta(p,o) = p \,.
\end{equation}
The inverse plane wave is expressed in terms of the antipode map $S :\mathbbm{R}^4 \to \mathbbm{R}^4$
\begin{equation}
\E[p] \, \E[S(p)] = \E[S(p)] \, \E[p] = \1 \,,
\end{equation}
satisfying the following properties, which, together with the previously-listed ones, mean that $(\mathbbm{R}^4,\Delta,S,\epsilon)$ close a commutative Hopf algebra:
\begin{equation}
\label{eq:antipode properties}
\Delta(p,S(p)) = \Delta(S(p), p) = o \,, \quad
S[\Delta(k,q)] = \Delta[S(q), S(k)] \,, \quad 
S(o) = o \,.
\end{equation}
The $*$ involution introduced in section \ref{intro} applied on $\E[p]$ yields the inverse exponential
\begin{equation}
\E^*[p] = \E[S(p)] \,. 
\end{equation}
The coproduct, counit and antipode codify the group product, unit element and inverse of the group $\mathcal{G}_\ell$ whose Lie algebra $C_\ell[\mathbbm{R}^{3,1}]$ is generated by the coordinate functions $\x^\mu$ and has commutation relations~\eqref{x_comm_rel}. Since the map $\Delta$ codifies the product of $\mathcal{G}_\ell$, one can use it to define left- and right- invariant vector fields
\begin{equation}
(X^\mu_\st{L})_\nu (p) = \left. \frac{\partial \Delta_\nu (p,q)}{\partial q_\mu} \right|_{q=o} \,,
\qquad
(X^\mu_\st{R})_\nu (p) = \left. \frac{\partial \Delta_\nu (q,p)}{\partial q_\mu} \right|_{q=o} \,,
\end{equation}
in terms of which we can define the left- and right- invariant Haar measure on $\mathcal{G}_\ell$:
\begin{equation}\label{Eq:Invariant_measures_definitions}
\begin{aligned}
&d\mu^\st{L} (p) = \left| \det  (X^\mu_\st{L})_\nu (p) \right|^{-1} \, d^4 p = 
 \left|  \det \frac{\partial \Delta_\nu (p,q)}{\partial q_\mu} \right|_{q=o}^{-1} \, d^4 p \,,
 \\
 &d\mu^\st{R}  (p) = \left| \det  (X^\mu_\st{R})_\nu (p) \right|^{-1} \, d^4 p = 
 \left|  \det \frac{\partial \Delta_\nu (q,p)}{\partial q_\mu} \right|_{q=o}^{-1} \, d^4 p \,.
\end{aligned}
\end{equation}
The left- and right- invariance is expressed as
\begin{equation}
\left. d\mu^\st{L} [\Delta(q,p)] \right|_{q\text{ fixed}}= d\mu^\st{L} (p) \,, \qquad \left. d\mu^\st{R} [\Delta(p,q)]  \right|_{q\text{ fixed}} = d\mu^\st{R} (p) \,,
\end{equation}
and the two measures are related by
\begin{equation}\label{LeftRightHaarMeasureRelation}
d\mu^\st{L} (p) =    d\mu^\st{R} [S(p)]  \,, \qquad 
d\mu^\st{R} (p) =    d\mu^\st{L} [S(p)]  \,. 
\end{equation}
The Fourier transform of a scalar function can thus be defined as 
\begin{equation}\label{LeftRightFourierTransformDefinition}
f(\x) = \int d \mu^\st{L} (p) \, \tilde{f}_\st{L} (p) \, \E[p] =\int d \mu^\st{R} (p) \, \tilde{f}_\st{R} (p) \, \E[p] \,, \qquad f(\x) \in C_\ell[\mathbbm{R}^{3,1}]\,,
\end{equation}
where $\tilde{f}_\st{L}$ and $\tilde{f}_\st{R}$ are two (commutative) functions, related to each other by
\begin{equation}\label{LeftRightFourierTransformRelation}
d \mu^\st{L} (p) \, \tilde{f}_\st{L} (p) = d \mu^\st{R} (p) \, \tilde{f}_\st{R} (p) 
~~ \Rightarrow ~~
\left| \det  (X^\mu_\st{L})_\nu (p) \right|^{-1} \, \tilde{f}_\st{L}(p)
=
\left| \det  (X^\mu_\st{R})_\nu (p) \right|^{-1} \, \tilde{f}_\st{R}(p) \,.
\end{equation}
For the case in which the group is unimodular, $d\mu^L(p)=d\mu^R(p)$ and thus $\tilde{f}_L$ and $\tilde{g}_L$ coincide.
Using $\E^*[p]=\E[S(p)]$ and acting with the involution on $f(\x)$, it is possible to show that
\begin{equation}\label{FourierTransformStarredFunction}
\widetilde{f^*}_\st{L}(p) = \overline{\tilde{f}_\st{R}[S(p)]}  
\,, 
\qquad
\widetilde{f^*}_\st{R}(p) = \overline{\tilde{f}_\st{L}[S(p)]}  \,.
\end{equation}
For a real function, the reality condition implies
\begin{equation}\label{FunctionRealityConditions}
\tilde{f}_\st{L}(p) = \overline{\tilde{f}_\st{R}[S(p)]}  
\,, \qquad \tilde{f}_\st{R}(p)= \overline{\tilde{f}_\st{L}[S(p)]} \, . 
\end{equation}

We can now introduce a notion of \textit{definite integral} (over all of spacetime) on noncommutative functions as a linear functional on $C_\ell[\mathbbm{R}^{3,1}]$:
\begin{equation}
 \int  (~.~) \, d^4\x : C_\ell[\mathbbm{R}^{3,1}] \to \mathbbm{C} \,,
\end{equation}
defined on the ordered exponentials as
\begin{equation}\label{NCintegralDefinition1}
\int  \, \E[p] \, d^4\x = \delta^{(4)}(p) \,,
\end{equation}
this implies
\begin{equation}\label{NCintegralDefinition2}
\int  \, f(\x) \, d^4\x =   \tilde{f}_\st{L}(o) =   \tilde{f}_\st{R}(o) \,, 
\end{equation}
which is possible because $(X^\mu_\st{L})_\nu (o) = (X^\mu_\st{L})_\nu (o) = \delta^\mu{}_\nu$, and therefore~\eqref{LeftRightFourierTransformRelation} implies $\tilde{f}_\st{L}(o) =   \tilde{f}_\st{R}(o)$.

The integral defined in~\eqref{NCintegralDefinition1} and~\eqref{NCintegralDefinition2} \textit{is not, in general, cyclic.} In fact

\begin{equation}\label{IntegralTwoFunctionsFourierTransform}
 \int \, f(\x) \, g(\x) \, d^4\x =\int \, \tilde{f}_\st{L}(p) \, \tilde{g}_\st{L}[S(p)] \,  d \mu^\st{L}(p) \neq \int \, \tilde{g}_\st{L}(p) \, \tilde{f}_\st{L}[S(p)] \,  d \mu^\st{L}(p)\,,
\end{equation}
which can only be equal when $d\mu^L(p)=d\mu^R(p)$. In \cite{Mercati:2024rzg} it was shown that this notion of integration is T-Poincaré invariant, in the sense that
\begin{equation}
     \int \, f(\x') \, d^4\x= \int \, f(\x) \, d^4\x \, .
\end{equation}

\subsection{Differential calculus and scalar field action}\label{Sec:DifferentialGeometry}
To construct the differential calculus, which we will need in order to build a T-Poincaré invariant action for the scalar field, we can introduce an exterior differential operator $d : C_\ell[\mathbbm{R}^{3,1}] \to \Gamma_\ell^1$ as in \cite{Woronowicz:1989}, such that:
\begin{equation}\label{Eq:Exterior_differential_def}
    d( \x^\mu) = d\x^\mu \,, \qquad d (\hat{f} \, \hat{g}) = d\hat{f} \, \hat{g} + \hat{f} \, d \hat{g} \,, \qquad d^2 = 0 \,,
\end{equation}
where $\hat{f}$ is shorthand for $f(\x)$.
The action of $d$ on generic functions can be described in terms of two linear operators  $\X^\mu, \, \Ch^\mu{}_\nu : C_\ell[\mathbbm{R}^{3,1}] \to C_\ell[\mathbbm{R}^{3,1}]$, which can be seen as elements of the translation subalgebra of $U_\ell[iso(3,1)]$, or, equivalently, as functions on momentum space. The indices of these operators are lowered and raised with the metric $\eta_{\mu\nu}$. The $\hat\xi_\mu$ operators are such that
\begin{equation} \label{ExteriorDifferentialExplicitForm}
d \hat{f} = i \, d \x^\mu \, (\X_\mu \triangleright \hat{f}) \,,
~~
\Delta(\X^\mu) = \X^\mu \otimes \1 + \Ch^\mu{}_\nu  \otimes \X^\nu \,,
~~
S(\X^\mu) = -(\Ch^{-1})^\mu{}_\nu  \X^\nu \,, \qquad \epsilon(\X_\mu) = 0 \,,
\end{equation}
while $\Ch^\mu{}_\nu$ is a grouplike element of $U_\ell[iso(3,1)]$:
\begin{equation}
\label{eq:chiproperties}
\Delta ( \Ch^\mu{}_\nu )= \Ch^\mu{}_\rho \otimes \Ch^\rho{}_\nu \,,
\qquad
S(\Ch^\mu{}_\nu) = (\Ch^{-1})^\mu{}_\nu \,, \qquad
\epsilon(\Ch^\mu{}_\nu) = \delta^\mu{}_\nu \,.
\end{equation}
More specifically, in \cite{Mercati:2024rzg} it is shown that $\Ch^\mu{}_\nu$ is actually a grouplike element of $U_\ell[so(3,1)]$, and thus satisfies 
\begin{equation}
\eta_{\mu\nu} \, \Ch^\mu{}_\rho(p) \, \Ch^\nu{}_\sigma(p) = \eta_{\rho\sigma} \,, \qquad
\eta^{\rho\sigma} \, \Ch^\mu{}_\rho(p) \, \Ch^\nu{}_\sigma(p) = \eta^{\mu\nu} \,.
\end{equation}
Using these properties and applying the exterior differential to a product of two functions, with the definition~\eqref{ExteriorDifferentialExplicitForm}, we get:
\begin{equation}
\label{ChiDefinitionFinal}
 d\x^\mu  \, \hat{f}  =
 (\Ch^\mu{}_\nu \triangleright \hat{f})   d\x^\nu \,, \qquad
 \hat{f} \,  d\x^\mu =  d\x^\nu  \, [(\Ch^{-1})^\mu{}_\nu  \triangleright \hat{f} ]\,.
\end{equation} 

The form of $\Ch^\mu{}_\nu$ can be read right off the commutation relations~\eqref{xdx_comm_rel}. In fact, the spacetime coordinates $\x^\mu$ act linearly, upon commutator, on $d \x^\mu$:
\begin{equation} \label{xdx_comm_rel_2}
 [ \x^{\mu} , d\x^{\nu} ]   
=
i \,  (K^\mu)^\nu{}_\rho   \,    d \x^\rho   \,.
\end{equation}
where $K^\mu$ are four Lorentz algebra matrices defined as
\begin{equation}\label{K_matrices_definition}
    (K^\mu)^\alpha{}_\beta = f^{\alpha\mu}{}_\beta \,.
\end{equation}
whose commutation relations can be found from \eqref{StructureConstants_constraints} and read
\begin{equation}\label{K_commutation_relations}
    [K^\mu , K^\nu] = 2 f^{[\mu\nu]}{}_\rho \, K^\rho \,,
\end{equation}
Therefore, an ordered exponential acts via adjoint action on $d \x^\mu$. If we consider an ordered exponential $\E[p]$, expressing the ordering choice terms of a factorization
\begin{equation}\label{Eq:OrderedExpGenericFactorization}
\E[p] = e^{i \, p^1_\mu \, \x^\mu} e^{i \, p^2_\mu \, \x^\mu} \dots e^{i \, p^n_\mu \, \x^\mu}\,,
\end{equation}
then the adjoint action of $\E[p]$ on $d \x^\mu$ is:
\begin{equation} \label{Eq:AdjointAction_on_dx}
\begin{aligned}
\E[p] \, d\x^{\mu} \, \E^*[p]   
=&
e^{i \, p^1_\mu \, \x^\mu} e^{i \, p^2_\mu \, \x^\mu} \dots e^{i \, p^n_\mu \, \x^\mu}
d\x^{\mu}
e^{-i \, p^n_\mu \, \x^\mu} e^{-i \, p^{n-1}_\mu \, \x^\mu} \dots e^{-i \, p^1_\mu \, \x^\mu}
\\
=&
e^{i \, p^1_\mu \, \x^\mu} e^{i \, p^2_\mu \, \x^\mu} \dots e^{i \, p^{n-1}_\mu \, \x^\mu}
\left(  e^{-p^{n}_\rho \, K^\rho}  \right)^{\mu}{}_{\nu} \, d \x^{\nu} \, e^{-i \, p^{n-1}_\mu \, \x^\mu} \dots e^{-i \, p^1_\mu \, \x^\mu}
\\
=&
\left(  e^{-p^{n}_\rho \, K^\rho}  \right)^{\mu}{}_{\nu} \, e^{i \, p^1_\mu \, \x^\mu} e^{i \, p^2_\mu \, \x^\mu} \dots e^{i \, p^{n-1}_\mu \, \x^\mu}
 d \x^{\nu} \, e^{-i \, p^{n-1}_\mu \, \x^\mu} \dots e^{-i \, p^1_\mu \, \x^\mu}
\\
&\vdots
\\
=&\left(  e^{-p^{n}_\rho \, K^\rho}  \right)^\mu{}_{\nu_1} \left(  e^{-p^{n-1}_\rho \, K^\rho}  \right)^{\nu_1}{}_{\nu_2} \dots \left(  e^{-p^{1}_\rho \, K^\rho}  \right)^{\nu_{n-1}}{}_{\nu_n} \, d \x^{\nu_n} \,.
\end{aligned}
\end{equation}
Notice that the order of the factors has been inverted, compared to Eq~\eqref{Eq:OrderedExpGenericFactorization}.
On the other hand, Eq.~\eqref{ChiDefinitionFinal} implies that
\begin{equation}
\label{eq:adjactiondx}
\E[p] \, d\x^{\mu} \, \E^*[p]   = d\x^\nu ( (\Ch^{-1})^\mu{}_\nu \triangleright \E[p] ) \E^*[p] \,,
\end{equation}
so we conclude that $\E[p] $ is an eigenfunction of the operator $(\Ch^{-1})^\mu{}_\nu$ with eigenvalue:
\begin{equation}\label{Eq:ConstructionChiMinusOne}
(\chi^{-1})^{\mu}{}_\nu(p) = \left(  e^{-p^{n}_\rho \, K^\rho}  \right)^\mu{}_{\nu_1} \left(  e^{-p^{n-1}_\rho \, K^\rho}  \right)^{\nu_1}{}_{\nu_2} \dots \left(  e^{-p^{1}_\rho \, K^\rho}  \right)^{\nu_{n-1}}{}_{\nu}  \,,
\end{equation}
which implies:
\begin{equation}
\label{eq:chiexpression}
\chi^\mu{}_\nu(p) = \left(: e^{ p_\rho \, K^\rho} : \right)^\mu{}_\nu \,,
\end{equation}
where the colons denote the same ordering as in Eq. \eqref{Eq:OrderedExpGenericFactorization}. Since $K^\mu$ are linear combinations of Lorentz matrices, $\chi^\mu{}_\nu(p)$ is also an $SO(3,1)$ matrix, \textit{i.e.}
\begin{equation}
\eta_{\mu\nu} \, \chi^\mu{}_\rho(p) \, \chi^\nu{}_\sigma(p) = \eta_{\rho\sigma} \,, \qquad
\eta^{\rho\sigma} \, \chi^\mu{}_\rho(p) \, \chi^\nu{}_\sigma(p) = \eta^{\mu\nu} \,.
\end{equation}
Analogously, by rexpressing \eqref{eq:adjactiondx} as
\begin{equation}
    \E[p]d\x^\mu\E^*[p]=E[p](\Ch^{\mu}{}_\nu\triangleright E[S(p)])d\x^\nu
\end{equation} it is further possible to show that $\chi^\mu{}_\nu(S(p))=(\chi^{-1})^\mu{}_\nu(p)$ in agreement with \eqref{eq:chiproperties}.

On the other hand, the functions $\xi_\mu(p)$ turn out to be a special coordinatization of momentum space. Indeed, in \cite{Mercati:2024rzg} it was shown that they satisfy the following relations,
\begin{equation}\label{Eq:Derivative_xi_in_zero}
\left. \frac{\partial \xi_\nu (q)}{\partial q_\alpha} \right|_{q=o}
=  \, \delta^\alpha{}_\nu \,, \qquad  \frac{\partial \xi_\mu (p)}{\partial p_\beta} 
\left. \frac{\partial \Delta_\beta(p,q)}{\partial q_\alpha}\right|_{q=o} =   \chi_\mu{}^\alpha (p)  \,, 
\end{equation}
which are obtained when differentiating both sides of $\xi_\mu[\Delta(p,q)] = \xi_\mu(p) + \chi_\mu{}^\nu (p) \, \xi_\nu (q)$ with respect to $q$. Using the definition of $d\mu^L(p)$ in \eqref{Eq:Invariant_measures_definitions} and the fact that $\chi^\mu{}_\nu$ is a $SO(3,1)$ matrix, one obtains
\begin{equation}
d\mu^\st{L} (p) 
 = d^4 \xi  \,.
\end{equation}
Therefore, momentum space coordinates $\xi_\mu$ are coordinates for which the left-invariant measure reduces to the commutative Poincaré invariant one.  We will refer to these as \textit{linear} momentum coordinates, for reasons which will be clear in the next section. 
In~\cite{Mercati:2024rzg}, the following formula is derived for the explicit calculation of $\xi_\mu(p)$:
\begin{equation}\label{Eq:ChiXiRepresentation}
\left(
\begin{array}{c|c}
\chi^\mu{}_\nu (p) & i \, \xi^\mu  (p)
\\
\hline
0 & 1
\end{array}
\right) =  ~ : \exp
\left(
\begin{array}{c|c}
  p_\alpha \, (K^\alpha)^\mu{}_\nu & i \, p^\mu 
\\
\hline
0 & 0
\end{array}
\right) : 
\,,
\end{equation}
where $p_\mu$ are the momentum-space coordinates associated to the ordering choice $: \, \cdot \, :$ for plane waves. As pointed out in~\cite{Mercati:2024rzg}, a generic ordering can be written as
\begin{equation}\label{Eq:FactorizationPlaneWave}
    \E[p] = ~ : \, \exp( i \, p_\mu \, \x^\mu)  \, : ~ = e^{i p^1_\mu \, \x^\mu} e^{i p^2_\mu \, \x^\mu} \dots e^{i p^n_\mu \, \x^\mu} \,,
\end{equation}
where, for example, the Weyl ordering is simply $n = 1$ and $p^1_\mu = p_\mu$, and the $\x^+$-to-the-right ordering has $n=2$ and $p^1_\mu = p_1 \, \delta^1_\mu+p_2 \, \delta^2_\mu+p_- \, \delta^-_\mu$, $p^2_\mu = p_+\delta^+_\mu$, where $p_\pm=\frac{p_0\pm p_3}{2}$.  In the representation~\eqref{Eq:ChiXiRepresentation}, we can then write the ordered exponential explicitly as
\begin{equation}\label{Eq:ChiXiRepresentation2}
\left(
\begin{array}{c|c}
\chi^\mu{}_\nu (p) & i \, \xi^\mu  (p)
\\
\hline
0 & 1
\end{array}
\right) = 
  \exp
\left(
\begin{array}{c|c}
  p^1 \cdot K  & i \, p^1
\\
\hline
0 & 0
\end{array}
\right)  \exp
\left(
\begin{array}{c|c}
  p^2 \cdot K  & i \, p^2
\\
\hline
0 & 0
\end{array}
\right) 
\dots \exp
\left(
\begin{array}{c|c}
  p^n \cdot K  & i \, p^n
\\
\hline
0 & 0
\end{array}
\right) 
\,,
\end{equation}
Consider a single exponential factor  in~\eqref{Eq:ChiXiRepresentation2}, we can split it as follows :
\begin{equation}\label{Eq:MatrixExponentialCalculation}
    \exp \left(
\begin{array}{c|c}
  p^i_\alpha \, (K^\alpha)^\mu{}_\nu & i \, (p^i)^\mu 
\\
\hline
0 & 0
\end{array}
\right)
=
\left(
\begin{array}{c|c}
 (\exp p_\alpha^i K^\alpha)^\mu{}_\nu   & i \, \Omega^\mu{}_\nu[p^i]\, (p^i)^\nu 
\\
\hline
0 & 1
\end{array}
\right)  \,,
\end{equation}
where $\Omega^\mu{}_\nu[q]$ is the $\phi_1$-function of $q_\rho K^\rho$:
\begin{equation}
\Omega^\mu{}_\nu[q] =   \left( \frac{e^{ q_\sigma \, K^\sigma} - I}{q_\rho \, K^\rho}\right)^\mu{}_\nu = \int_0^1  \left(e^{(1- s) \, q_\sigma \, K^\sigma} \right)^\mu{}_\nu \, ds \,.
\end{equation}
% Then, multiplying all matrices, one gets:
% \begin{equation}\label{Eq:Calculation_xi}
%     \begin{aligned}
% \chi^\mu{}_\nu (p)  =& \left( e^{p^1 \cdot K}e^{p^2 \cdot K}  \dots e^{p^n \cdot K}\right)^\mu{}_\nu \,,
% \\
%  \xi^\mu  (p) =& \Omega^\mu{}_\nu[p^1]\, (p^1)^\nu
% +\left( e^{p^1 \cdot K}\right)^\mu{}_{\rho_1} \bigg{(} \Omega^{\rho_1}{}_\nu[p^2]\, (p^2)^\nu
% +\left(e^{p^2 \cdot K}\right)^{\rho_1}{}_{\rho_2}  \bigg{(}\Omega^{\rho_2}{}_\nu[p^3]\, (p^3)^\nu
% + \dots 
% \\
% &+ \left(e^{p^{n-1} \cdot K}\right)^{\rho_{n-2}}{}_{\rho_{n-1}}  \Omega^{\rho_{n-1}}{}_\nu[p^n]\, (p^n)^\nu
% \bigg{)} \dots \bigg{)} \bigg{)}\,.
%     \end{aligned}
% \end{equation}
Then, multiplying all terms in~\eqref{Eq:ChiXiRepresentation2}, one gets an expression that depends on the following matrices:
\begin{equation}
\label{eq:Umat}
   (U_0)^\mu{}_\nu = \delta^\mu{}_\nu \,, \quad 
    (U_1)^\mu{}_\nu  = \left( e^{p^1 \cdot K} \right)^\mu{}_\nu \,, ~~\dots,~~
    (U_n)^\mu{}_\nu  = \left( e^{p^1 \cdot K}e^{p^2 \cdot K}  \dots e^{p^n\cdot K}\right)^\mu{}_\nu \,;
\end{equation}
which is
\begin{equation}\label{Eq:Calculation_xi}
\chi^\mu{}_\nu (p)  = U_n \,,
\qquad
 \xi^\mu  (p) = \sum_{j=1}^{n} (U_{j-1})^\mu{}_\rho \,  \Omega^\rho{}_\nu [p^j]\, (p^j)^\nu\,.
\end{equation}

To complete the description of the differential calculus, in \cite{Mercati:2024rzg} it was shown that the wedge product $\wedge$ and the spaces of $n$-forms $\Gamma^n_\ell$ can be introduced as well as left- and right-$C_\ell[\mathbbm{R}^{3,1}]$-modules, and the basis $n$-forms close an undeformed exterior algebra. The differential can be generalized to a map $d: \Gamma^n_\ell \to \Gamma^{n+1}_\ell$, satisfying the standard graded Leibniz rule. Finally, the Hodge star will be a left- and right-$C_\ell[\mathbbm{R}^{3,1}]$-linear map, with undeformed action on the basis $n$-forms.

All the differential geometry structures presented so far allow to write the most general action, invariant under T-Poincaré transformations, for a free (real or complex) scalar field (for further details refer to~\cite{Mercati:2024rzg}). In the present paper we will be interested in the case of a real scalar field, for which the most general action for a free field, in Fourier transform, takes the form 
\begin{equation}\label{Eq:ScalarAction_FourierTransform}
    S = {\frac 1 2} \int d^4 \xi \, \tilde{\phi}_\st{L} (\xi) \tilde{\phi}_\st{L}[S(\xi)] \, \mathcal{C}(\xi) = {\frac 1 2} \int d^4 \xi \, \left| \tilde{\phi}_\st{L}(\xi) \right|^2 \, \mathcal{C}(\xi)  \,,
\end{equation}
where we used the reality condition $\overline{\tilde{\phi}(\xi)}=\tilde{\phi}[S(\xi)]$. Notice that the antipode of $\xi$ is nontrivial, and does not in general coincide with $-\xi$. This action has been conveniently written in terms of linear momentum coordinates and the function $\mathcal{C}(\xi)$ is just (a function of) the undeformed Casimir element
\begin{equation}
\mathcal{C}(\xi) =\eta^{\mu\nu}\xi_\mu\xi_\nu + m^2=\eta^{\mu\nu}S(\xi)_\mu S(\xi)_\nu + m^2 \,,
\end{equation}
where the second equality can be derived from \eqref{ExteriorDifferentialExplicitForm}. The action~\eqref{Eq:ScalarAction_FourierTransform} is perfectly equivalent to that of a commutative field theory, written in Fourier transform. We can extremize it by varying with respect to $\tilde{\phi}_\st{L}(k)$ and its complex conjugate, as we would do in the commutative case:
\begin{equation}
\delta \mathcal{S}  = \int  \mathcal{C}(\xi) \, \left(\delta  \overline{\tilde{\phi}_\st{L} (\xi)} \, \tilde{\phi}_\st{L}(\xi) +  \overline{\tilde{\phi}_\st{L} (\xi)} \,\delta  \tilde{\phi}_\st{L}(\xi)  \right) \, d^4\xi \,,
\end{equation}
yielding the equations of motion
\begin{equation}\label{Eq:ScalarEOM}
\mathcal{C}(\xi) \, \overline{\tilde{\phi}_\st{L} (\xi)}
= \mathcal{C}(\xi) \, \tilde{\phi}_\st{L} (\xi) = 0\,,
\end{equation}
which are simply solved by requiring that $\tilde{\phi}_\st{L} (\xi)$ is supported on the region $\mathcal{C}(\xi)=0$, the on-shell condition.  The extremizing field configurations take the form
\begin{equation}\label{Eq:OnShellScalarFields}
 \tilde{\phi}_\st{L}(\xi)
 = g_L(\xi) \delta \left[ \mathcal{C}(\xi) \, \right]  \,,
\qquad
 \overline{\tilde{\phi}_\st{L} (\xi)} \,,
  =  \overline{g_\st{L}(\xi)} \, \delta\left[ \mathcal{C}(\xi) \, \right]   \,,
\end{equation}
for some momentum space function $g_\st{L}(\xi)$.
In \cite{Mercati:2024rzg} it was shown that by extremizing the action in configuration space, one obtains the Fourier transformed result of \eqref{Eq:ScalarEOM}, namely
\begin{equation}
\label{eq:NCKG}
    \square \triangleright \phi(\x)-m^2\phi(\x)=0 \, .
\end{equation}

\subsection{Properties of plane waves in the braided tensor product algebra}
\label{subsec:theorems}
Before moving on to the main results of the paper, we show some properties concerning coordinate differences, elements of $\mathcal{M}^{N-1}\subset C_\ell[\mathbbm{R}^{3,1}]^{\otimes_\ell N}$ and linear momentum coordinates $\xi_\mu$ which are not reported in \cite{Mercati:2024rzg}. These are valid for any T-Minkowski model and prove useful in deriving the results of the next section.

First, we show that
\begin{equation}
\label{eq:th1}
    \E_a[p]h(\x_b-\x_c)=h\left[(\chi^{-1})^\mu{}_\nu(p)(\x_b^\nu-\x_c^\nu)\right]\E_a[p] \, ,
\end{equation}
where $h$ is a generic function depending on the coordinate difference $\x_b-\x_c$, with $a,b,c=1,...,N$. This tells us that when commuting a generic plane wave with a function that depends solely on coordinate differences, the effect is to introduce the (momentum-dependent) $SO(3,1)$ matrix $\chi^\mu{}_\nu$, defined in \eqref{eq:chiexpression},  which transforms the coordinate difference with a Lorentz transformation.
To show this, recall equation \eqref{eq:commcoorddiff}:
$$
    [\x_a^\mu,\x_b^\nu-\x_c^\nu]=i \, f^{\nu\mu}_\rho(\x_b^\rho-\x_c^\rho) \, , 
$$
and the fact that $f^{\nu\mu}_\rho=(K^\mu)^\nu_\rho$ where $K\in\mathfrak{so}(3,1)$. 
Given that the commutation relations between $\x_a^\mu$ and $\x_b^\nu-\x_c^\nu$ have the same structure as those between $\x_a^\mu$ and $d\x^\nu$, we can reason as in \eqref{Eq:AdjointAction_on_dx} and conclude that
\begin{equation}\label{Adjoint_action_Exp_Coord_diff}
   \E_a[p](\x_b-\x_c)^\mu\E_a^*[p]= (\chi^{-1})^\mu{}_\nu  (p) (\x_b^\nu-\x_c^\nu) \,.
\end{equation}
By expanding a generic function $h$ in a formal power series of the coordinate difference, one can thus prove that 
\begin{equation}
\label{eq:fconj}
    \E_a[p]h(\x_b-\x_c)\E_a^*[p]=h\left[\E_a[p](\x_b-\x_c)\E_a^*[p]\right]
    =h\left[(\chi^{-1})^\mu{}_\nu  (p) (\x_b^\nu-\x_c^\nu)\right]\, . 
\end{equation}
which is equivalent to ~\eqref{eq:th1}.

Consider now an ordered plane wave of point $a$. The analogue of formula~\eqref{Eq:FactorizationPlaneWave} is
\begin{equation}
    \E_a[p] = ~ : \, \exp( i \, p_\mu \, \x^\mu_a)  \, : ~ = e^{i p^1_\mu \, \x^\mu_a} e^{i p^2_\mu \, \x^\mu_a} \dots e^{i p^n_\mu \, \x^\mu_a} \,,
\end{equation}
and we can separate the center-of-mass coordinates $\x^\mu_{cm}$ from the relative ones, $\y^\mu_a$ in each exponential factor:
\begin{equation}
    e^{i p^i_\mu \, \x^\mu_a}  = 
    e^{i p^i_\mu (\x^\mu_{cm}+\y^\mu_a)}  \,.
\end{equation}
From the commutation relations~\eqref{eq:comcommrels} and the semidirect product structure of the Lie algebra one sees that the adjoint action of the center-of-mass coordinates $\x^\mu_{cm}$ on the relative coordinates $\y^\mu_a$  is identical to that of the coordinates $\x^\mu$ on the differentials $d\x^\mu$, Eq.~\eqref{xdx_comm_rel_2}, involving the action of the matrix $K^\mu$ on $\y_a^\nu$. Then, one can split the last exponential analogously to Eq.~\eqref{Eq:MatrixExponentialCalculation}:
\begin{equation}
    e^{i p^i_\mu (\x^\mu_{cm}+\y^\mu_a)}  
    =
    e^{i \eta_{\rho\mu} \, \Omega^\mu{}_\nu[p^i](p^i)^\nu \, \y^\rho_a}  \,
    e^{i p^i_\mu \, \x^\mu_{cm}}  \,.
\end{equation}
Putting in sequence all the factors of the ordered exponential, and reordering the $\exp(i p^i_\mu \, \x^\mu_{cm})$ factors to the right through Eq.~\eqref{eq:comcommrels},  one ends up with the expression 
\begin{equation}
    \E_a[\p]=e^{i \alpha_\mu(p)\y_a^\mu}  : \, e^{i p_\mu \, \x^\mu_{cm}} \, : \, ,
\end{equation}
where 
\begin{equation}
\alpha_\mu(p)=\sum_{j=1}^n \eta_{\sigma\rho}(U_{j-1}^{-1})^\sigma{}_\mu \Omega^\rho{}_\nu[p^j](p^j)^\nu \, ,
\end{equation}
and the $U_j^{-1}$ matrices are just the inverse of the ones defined in \eqref{eq:Umat}. Using the fact that they are Lorentz matrices, we can use the relation $\eta_{\rho\sigma}(U_{j-1}^{-1})^{\sigma}{}_\mu=\eta_{\mu\sigma}(U_{j-1})^\sigma{}_\rho$ to write
\begin{equation}
    \alpha_\mu(p)=\sum_{j=1}^n \eta_{\mu\rho}(U_{j-1})^\rho{}_\sigma \Omega^\sigma{}_\nu[p^j](p^j)^\nu=\xi_\mu(p) \, ,
\end{equation}
which simply follows from the definition of $\xi^\mu(p)$ from Eq.~\eqref{Eq:Calculation_xi}. We conclude that
\begin{equation}
\label{eq:pwsplitting}
    \E_a[p] = e^{i \xi_\mu(p) \, \y^\mu_a}  \,
    : \, e^{i p_\mu \, \x^\mu_{cm}} \, : \,.
\end{equation}
As a corollary, we have that $\E_a[p] \E_b^*[p]$, $\E_b^*[p] \E_a[p] \in   \mathcal{M}^{N-1}$:
\begin{equation}
\label{eq:pwprod}
    \E_a[p]\, \E_b^*[p] = e^{i \,\xi_\mu(p)(\x_a-\x_b)^\mu} 
    \qquad
    \E_b^*[p]\, \E_a[p] = e^{i \,\xi_\mu[S(p)](\x_a-\x_b)^\mu}=e^{iS[\xi_\mu(p)](\x_a-\x_b)^\mu} \,.
\end{equation}
We are now in a position to write T-Poincaré-invariant N-point functions, which we will show to be elements of $\mathcal{M}^{N-1}$. This is a generalization of a result first derived in \cite{Lizzi:2021rlb} in the context of the lightlike $\kappa$-Minkowski spacetime.
Consider a generic function living in the braided tensor product algebra $C_\ell[\mathbbm{R}^{3,1}]^{\otimes_\ell N}$, whose Fourier transform we can write as\footnote{The N-point algebra $C_\ell[\mathbbm{R}^{3,1}]^{\otimes_\ell N}$ satisfies the  Poincaré--Birkhoff--Witt property,  therefore
any polynomial element of the algebra can be written uniquely as an \textit{ordered} polynomial. The ordering chosen in~\eqref{eq:genericfunction} is the following: $x^\mu_a$ is to the left of $x^\nu_b$ if $a<b$, and if $a=b$ the ordering choice of $\E_a$ between the four coordinates of the same point is followed. $C_\ell[\mathbbm{R}^{3,1}]^{\otimes_\ell N}$ is a Lie algebra, and therefore it can be treated as a generalized Weyl system in the sense of~\cite{Agostini:2002de}, and introduce a basis of ordered plane waves and a notion of noncommutative Fourier transform. A generic Fourier-transformable function can then be written as in Eq.~\eqref{eq:genericfunction}, where $f(p_a)$ is a commutative Schwarz function.}
\begin{equation}
\label{eq:genericfunction}
    f(\x_1,\dots,\x_N)=\int d\mu_L(p_1)\dots d\mu_L(p_N)\tilde{f}_L(p_a)\E_1[p_1]\cdots \E_N[p_N] \, ,
\end{equation}
where the subscripts and the index $a$, running from $1$ to $N$, distinguish the four-momenta parametrizing the plane waves of each copy of $C_\ell[\mathbbm{R}^{3,1}]^{\otimes_\ell N}$.
The plane waves $\E_a[p_a]$, written using any ordering prescription, can be conveniently expressed in terms of center of mass coordinates and coordinate differences, according to \eqref{eq:pwsplitting}:

\begin{equation}
    \E_a[p_a] =:e^{ip_{a,\mu}(\y_a^\mu+\x_{cm}^\mu)}:= e^{i \xi_\mu(p_a) \, \y^\mu_a}  \,
    : \, e^{i p_{a,\mu} \, \x^\mu_{cm}} \, : \,.
\end{equation}
Then, all the terms depedning on $\x_{cm}$ can be commuted to the right, using the following relation
\begin{equation}
    : \, e^{i p_{a,\mu} \, \x^\mu_{cm}}  \, : e^{i \xi_\mu(p_b) \, \y^\mu_b}  \,= e^{i\xi_\mu(p_b)(\chi^{-1}(p_a))^\mu{}_\nu \y_b^\nu}  : \, e^{i p_{a,\mu} \, \x^\mu_{cm}}  \, : \, ,
\end{equation}
which can be obtained analogously to relation \eqref{Adjoint_action_Exp_Coord_diff} given the commutation relations \eqref{eq:comcommrels}. In turn, from \eqref{eq:comcommrels} we derive another useful property 
 \begin{equation}
     :e^{ip_{a,\mu} \x_{cm}^\mu}::e^{ip_{b,\mu} \x_{cm}^\mu}:=:e^{i\Delta_\mu[p_a,p_b] \x^\mu_{cm}}\, : \, ,
 \end{equation}
 where $\Delta_\mu[p_a,p_b]$ is defined by \eqref{ProductBetweenExponentials}.
 Therefore, the generic function in \eqref{eq:genericfunction} can be rewritten as
\begin{equation}
\label{eq:genericfunctionv2}
\begin{aligned}
    f(\x_1,\dots,\x_N)=&\int d\mu_L(p_1)\dots d\mu_L(p_N)\tilde{f}_L(p_a)e^{i \xi_\mu(p_1) \, \y^\mu_1}  \,
    : \, e^{i p_{1,\mu} \, \x^\mu_{cm}} \, :\dots e^{i \xi_\mu(p_N) \, \y^\mu_N}  \,
    : \, e^{i p_{N,\mu} \, \x^\mu_{cm}} \, : \, =\\
    =&\int d\mu_L(p_1)\dots d\mu_L(p_N)\Big[\tilde{f}_L(p_a)e^{i\xi_\mu(p_1)\y_1^\mu}e^{i(\chi^{-1}(p_1) \xi(p_2))_\mu \y_2^\mu}\dots\times \\
    \times & e^{i(\chi^{-1}(p_1)\dots \chi^{-1}(p_{N-1})\cdot \xi(p_N))_\mu \y_N^\mu }:e^{i(\Delta[p_1,\dots,p_N])_\mu \x^\mu_{cm}}: \Big] \, .
\end{aligned}
\end{equation}
The only way that such a function can be invariant under translations is if the dependence on $\x_{cm}$ disappears, which is achieved if and only if $\Delta[p_1,\dots,p_N]=0$. Our $N$-point function \eqref{eq:genericfunctionv2} can be further simplified performing the change of variables
\begin{equation}
\begin{aligned}
    &\xi_{1}=\xi(p_1)\\
    &\xi_{2}=\xi(p_2) \cdot  (\chi^{-1}(p_1))\,,\\
    &\vdots\\
    &\xi_{N}=\xi(p_N) \cdot (\chi^{-1}(p_{N-1}))\cdot \dots  \, \cdot (\chi^{-1}(p_1)) \, .
    \end{aligned}
\end{equation}
Recalling that $d\mu_L(p_a)=d^4\xi_a$ and that 
$\chi^{-1}(p_a)$ are Lorentz matrices, \eqref{eq:genericfunctionv2} can be finally expressed as 
\begin{equation}
\label{eq:genericfunctionv3}
    f(\x_1,\dots,\x_N)=\int d^4\xi_1\dots d^4\xi_N\tilde{f}_L(\xi_1,\dots\xi_N)e^{i\xi_{1,\mu}\y_1^\mu}\cdots e^{i\xi_{N,\mu}\y_N^\mu}
\end{equation}
Then, invariance under the remaining Lorentz sector of the transformation is guaranteed with particular choices of $f$, as we will see below.

Recall that the algebra of functions on T-Minkowski spacetime, $C_\ell[\mathbbm{R}^{3,1}]$ and its braided tensor product $C_\ell[\mathbbm{R}^{3,1}]^{\otimes_\ell N}$, can be expressed in any basis of ordered plane waves, for whatever choice of ordering. These bases are all related to each other by a (possibly nonlinear) map acting on the momenta $p_\mu$ labeling the plane waves $\E_a[p]$. The Lie group structure of momentum space actually makes these change of basis more than continuous functions of the momenta: they are \textit{diffeomorphisms} in momentum space~\cite{Mercati:2024rzg}. This is a natural invariance of the model which should be a symmetry of the physical theories built upon it: in our calculations we should be able to expand our fields in any basis of ordered plane waves, and the physical predictions of the theory should be identical. So we would like to have a form of diffeomorphism invariance in momentum space. This goes beyond mere changes of ordering: any coordinate system in momentum space should be equivalent. In particular, it is convenient to use $\xi_\mu$ as coordinates on momentum space; at least in unimodular models, they are a good coordinate system because the map $p_\mu \to \xi_\mu(p)$ is a bijection\footnote{An example of a non-unimodular model in which this is not the case can be found in Ref. \cite{Fabiano:2023xke}, where covariant QFT on the lightlike $\kappa$-Minkowski spacetime is developed in 1+1 dimensions. The map from standard momentum coordinates to linear momentum coordinates in that case reads $\xi_-=p_-, \xi_+=\frac{1}{2}(e^{2p_+}-1)$, which is clearly not a bijection.} (see \cref{app:Rmatrices}). This coordinate system is not necessarily associated to any particular ordering (although it may be), but it is convenient for a very good reason: \textit{in these coordinates, momenta transform linearly under Lorentz transformations.} Let us prove this property.

If we act with a T-Poincaré transformation \eqref{Coaction_spacetime_coordinates} on the plane wave product in \eqref{eq:pwprod}, we obtain
\begin{equation}
\label{eq:xitransf1}
    \left(\E_a[p] \, \E_b^{*}[p]\right)'=\left(e^{i\xi_\mu(p)(\x_a-\x_b)^\mu}\right)'=e^{i\xi_\mu(p) \, \La^\mu{}_\nu\otimes (\x_a-\x_b)^\nu}
    \,.
\end{equation}
On the other hand, the T-Poincaré coaction is a homomorphism for the product of $C_\ell[\mathbbm{R}^{3,1}]^{\otimes_\ell N}$, so the  left hand side of this equation can also be written as $\E_a'[\xi]\E_b^{'*}[\xi]$. Recalling that a T-Poincaré transformed plane wave can always be written as~\cite{Mercati:2024rzg}:
\begin{equation}
    \E_a'[p] = ~ :\, e^{i \, \lambda_\mu(p,\La)\otimes \x^\mu} \, : \, : \, e^{i \, p_\mu \A^\mu\otimes 1} \,: \,,
\end{equation}
where $: \, . \, :$ is the chosen ordering for $\x^\mu$'s (applied in the same way to $\A^\mu$'s ), and $\lambda_\mu(p,\La)$ is a (in general nonlinear) representation of the Lorentz group as maps on momentum space. 
Then, the product $\E_a'[p]\E_b^{'*}[p]$ will be independent of $\A^\mu$, and thus
\begin{equation}
\label{eq:xitransf2}
 \E_a'[p]\E_b^{'*}[p] = e^{i\xi_\mu[\lambda(p,\La)]\otimes(\x_a-\x_b)^\mu} = e^{i\xi'_\mu \otimes(\x_a-\x_b)^\mu} \,.
\end{equation}
Comparing \eqref{eq:xitransf1} and \eqref{eq:xitransf2}, we obtain
\begin{equation}
    \label{eq:xilintrasnf}
    \xi'_\mu =
    \xi_\mu[\lambda(p,\La)]  = \La^{\nu}{}_\mu  \, \xi_\nu(p) \,,
\end{equation}
meaning that $\xi_\mu$ transforms linearly under Lorentz transformations\footnote{A similar argument, using the opposite order for plane waves, $\E_b^*[p] \,  \E_a[p] $, reveals that $\E_b^{'*}[p]\E'_a[p] = e^{i S(\xi)_\mu[\lambda(p,\La)]\otimes(\x_b-\x_a)^\mu}$ and consequently $S(\xi)_\mu[\lambda(p,\La)] = \xi_\mu[\lambda(p,\La)]\chi^\mu{}_\nu[\lambda(p,\La)] =  (\La \triangleleft p)^\nu{}_\mu \, S(\xi)_\nu(p)$, where the so-called backreaction $\La\triangleleft p$ was defined in \cite{Mercati:2024rzg}.}. 

Using this result we can also deduce the infintesimal action of the T-Poincaré group on arbitrary noncommutative functions. Let us start by the T-Poincaré transformation of a plane wave, expanded as a power series in the group parameters, around $\A^\mu=0$ and $\La^\mu{}_\nu=\delta^\mu_\nu\1$ . Using \eqref{eq:xilintrasnf}, which can be inverted as $\lambda_\mu(p(\xi),\La)=p_\mu(\xi_\nu \La^\nu_{\ \rho})$, we obtain, up to first order in the group parameters
\begin{equation}
    \label{eq:pwexpansion}\E_a'[p(\xi)]\simeq\1\otimes \E_a[p(\xi)]+i\A^\mu p_\mu(\xi)\otimes\E_a[p(\xi)]+i\xi_\mu(\La^\mu{}_\nu-\delta^\mu_{\ \nu} \,\1)\otimes\frac{\partial\E[p(\xi)]}{\partial\xi_\nu} \, ,
\end{equation}
where $p_\mu(\xi)$ is the (in general) non-linear relationship between $p_\mu$ and linear momentum coordinates which can be found using relation \eqref{Eq:Calculation_xi}.
Defining
\begin{equation}
    \La^\mu{}_\nu=\delta^\mu_{\ \nu} \,\1 +\hat \omega^{\mu\rho}\eta_{\rho\nu}+\mathcal{O}(\hat{\omega}^2) \,,
\end{equation}
we can rewrite \eqref{eq:pwexpansion} in terms of a left action of infinitesimal generators $\hat{P}_\mu,\M_{\rho\sigma}$ 
as
\begin{equation}
\label{eq:leftaction}
    \E_a'[p(\xi)]\simeq \1\otimes \E_a[p(\xi)]+i\,\A^\mu\otimes \left[\hat P_\mu\triangleright \E_a[p(\xi)]\right]+i \, \hat \omega^{\rho\sigma}\otimes \left[\M_{\rho\sigma}\triangleright\E_a[p(\xi)]\right]\,.
\end{equation}
Comparing \eqref{eq:pwexpansion} and \eqref{eq:leftaction}, we obtain 
\begin{equation}
\label{eq:genaction}
    \hat P_\mu\triangleright \E_a[p(\xi)]=p_\mu(\xi) \E_a[p(\xi)] \qquad \M_{\rho\sigma}\triangleright \E_a[p(\xi)]=(\xi_\rho\eta_{\sigma\mu}-\xi_\sigma\eta_{\mu\rho})\frac{\partial \E_a[p(\xi)]}{\partial \xi_\mu}
\end{equation}
Expressions \eqref{eq:genaction} can be used to compute the left action of these generators on a noncommutative function:
\begin{equation}
\label{eq:actiononfunct}
\begin{aligned}
    \hat P_\mu\triangleright f(\x_a)=&\int d^4\xi \,\tilde f_\st{L}(\xi) p_\mu(\xi) \E_a[p(\xi)] \,,\\
    \M_{\rho\sigma}\triangleright f(\x_a)=&\int d^4\xi \tilde f_\st{L}(\xi) (\xi_\rho\eta_{\sigma\mu}-\xi_\sigma\eta_{\mu\rho})\frac{\partial \E_a[p(\xi)]}{\partial \xi_\mu} \,.
    \end{aligned}
\end{equation}
Notice that the action of the Lorentz generators $\hat M_{\rho\sigma}$ on a generic function is linear.
Integrating by parts the second relation in \eqref{eq:actiononfunct}, it is possible to derive the action of the commutator $[\M_{\rho\sigma},\hat P_\mu]$ on a generic noncommutative function: 
\begin{equation}
\label{eq:commaction}
    [\M_{\rho\sigma},\hat P_\mu]\triangleright f(\x_a)=\int d^4\xi\, \tilde{f}_\st{L}(\xi) (\xi_\rho\eta_{\sigma\nu}-\xi_\sigma\eta_{\nu\rho})\frac{\partial p_\mu(\xi)}{\partial \xi_\nu}  \E_a[p(\xi)] \,,
\end{equation}
By defining the operator $\X_\mu(\hat P)$, such that 
\begin{equation}
\label{eq:linmomop}
    \X_\mu \triangleright \E_a[p(\xi)]=\xi_\mu(p) \E_a[p(\xi)] \,,
\end{equation}
by means of \eqref{eq:commaction}, it is easy to show that
\begin{equation}
\label{eq:linPoinc}
    [\hat M_{\rho\sigma},\X_\mu]=\X_\rho\eta_{\sigma\mu}-\X_\sigma\eta_{\mu\rho} \,,
\end{equation}
namely, the operators $\hat{M}_{\rho\sigma}$ and $\hat{\xi}_\mu$ close the standard Poincaré algebra. This was to be expected, given the linear transformation properties of $\xi_\mu$ reported in \eqref{eq:xilintrasnf}.

Given the useful properties of linear momentum coordinates obtained in this subsection, we will make ample use of these variables to perform computations in the remainder of the manuscript. Thus, to ease the notation, we will denote plane waves $\E[p(\xi)]$ simply as $\E[\xi]$. With respect to plane waves $\E[p]$ parametrized by the ``standard'' momentum $p$, we pay the price of having (in general) nonlinear functions of the momenta $p(\xi)$ appearing in the exponent, but gain the great advantage of being able to perform standard Poincaré transformations on plane waves $\E[\xi]$, thanks to relation \eqref{eq:linPoinc}. Most importantly, this is just a calculation device: our result will not depend on the choice of coordinates on momentum space used to label plane waves and Fourier transforms.

In concluding this subsection, let us focus on T-Poincaré-invariant two-point functions, written in terms of $\E[\xi]$ plane waves, which will be pivotal in developing the quantization procedure of \Cref{sec:ncqft}. From the result \eqref{eq:genericfunctionv3}, we know that such a distribution must depend on a coordinate difference, so there are two natural ways to write it: 
\begin{equation}
\label{eq:2pointdistr}
\begin{aligned}
    F_\st{L}(\x_a-\x_b) =& \int d^4\xi \, h_L(\xi) \,  \E_a[\xi]\, \E_b^*[\xi]=\int d^4\xi\, h_L(\xi)\, e^{i\xi_\mu(\x_a-\x_b)^\mu}\\
    F_\st{R}(\x_a-\x_b)=&\int d^4S(\xi) \,h_R(\xi) \E_a^*[\xi]E_b[\xi]=\int d^4S(\xi)\,h_R(\xi)e^{iS_\mu(\xi)(\x_a-\x_b)^\mu}=\\
    =&\int d^4\xi\,  \left|\det\frac{\partial S(\xi)}{\partial \xi}\right|h_R(S(\xi))e^{i\xi_\mu(\x_a-\x_b)^\mu}=\int d^4\xi\,  h'_R(S(\xi))e^{i\xi_\mu(\x_a-\x_b)^\mu} \, ,
\end{aligned}
\end{equation}
where going from the second to the third line, we have performed the change of variables $S(\xi)\rightarrow \xi $ and absorbed the determinant of the Jacobian in the definition of $h'_R(S(\xi))$. Eq. \eqref{eq:2pointdistr} clearly shows that the two choices are actually equivalent. Given the transformation rules of $\xi_\mu$ in \eqref{eq:xilintrasnf}, $d^4\xi$ is invariant under T-Poincaré transformations, so the invariance of the two-point distribution is guaranteed only by choosing $h_{L,R}$ as Lorentz-invariant functions of momentum space. As a result, they may be functions of the mass Casimir $\xi_\mu\xi^\mu=S(\xi_\mu)S(\xi^\mu)$ and, in the case where $\xi$ is timelike, they may also depend on the sign of $\xi_0$.

\subsection{Universal R-matrix}
\label{sec:rmat}
All T-Minkowski models are equipped with an R-matrix which governs the commutation relations of the braided tensor product algebra. In \cite{Mercati:2023apu,Mercati:2024rzg}, only a finite-dimensional representation for the R-matrices was provided, while here we utilize an infinite-dimensional one, which we will use to obtain the main results of our paper in the next section. Our starting point is the result \cite{tolstoy2007twistedquantumdeformationslorentz}, where Drinfel'd twists~\cite{Drinfeld_thm} for all T-Minkowski models are calculated. From a twist element $\mathcal{F}$, one can define an R-matrix as
\begin{equation}
\label{eq:Rmatdef}
   \mathcal{R} = \mathcal{F}_{21}^{-1} \, \mathcal{F}_{12} \,,
\end{equation}
where $\mathcal{F}_{21}=\tau\circ \mathcal{F}_{12}$ and $\tau$ is the flip operator.
According to the list of twist elements found in \cite{tolstoy2007twistedquantumdeformationslorentz} and the definition \eqref{eq:Rmatdef}, for all unimodular T-Minkowski models with the cocycle set to zero, the R-matrix can be written in the form
\begin{equation}
\label{eq:Rmatgenform}
    \mathcal{R}=  \mathcal{R}_n \,  \mathcal{R}_{n-1} \, \dots \, \mathcal{R}_1  \,, \qquad
\mathcal{R}_i =    
    \exp\left[ \,\gamma^{\mu\nu\rho}_i \, P_{\mu}(\hat \xi)\otimes \M_{\nu\rho}\right]\exp\left[-\, \gamma^{\mu\nu\rho}_i \,\M_{\nu\rho}\otimes  P_{\mu} (\hat \xi)\right] \,,
\end{equation} 
where $\gamma^{\mu\nu\rho}_i$ are numerical coefficients, and $P_\mu(\hat \xi)$ is a (in general) nonlinear function of operators $\hat \xi$ with the same functional dependence of $p(\xi)$ introduced in the previous subsection. We will use the R-matrix to swap the product of plane waves~\cite{majid_1995}:
\begin{equation}
\label{eq:pwex}
\E_b[\eta]\E_a[\xi]=\mu\circ \mathcal{R} \triangleright \E_a[\xi]\otimes \E_b[\eta] \, ,
\end{equation}
where $\mu$ indicates the product in $C_\ell[\mathbbm{R}^{3,1}]^{\otimes_\ell N}$. To illustrate how \eqref{eq:pwex} works, let us focus on one of the factors in \eqref{eq:Rmatgenform} acting on the tensor product of two plane waves, $\E_a[\xi]\otimes\E_b[\eta]$. Using \eqref{eq:genaction} and \eqref{eq:linmomop}, one can see that
\begin{equation}
    \begin{aligned}
        \mathcal{R}_i \triangleright \E_a[\xi]\otimes \E_b[\eta] = &\exp\left[ \,\gamma^{\mu\nu\rho}_i \, P_{\mu}(\hat \xi)\otimes \M_{\nu\rho}\right]\exp\left[-\,\gamma^{\mu\nu\rho}_i \,\M_{\nu\rho}\otimes  P_{\mu} (\hat \xi)\right] \triangleright \E_a[\xi]\otimes \E_b[\eta]=
\\
        =&\exp\left[ \,\gamma^{\mu\nu\rho}_i \, P_{\mu}(\hat \xi)\otimes \M_{\nu\rho}\right]\exp\left[-\,\gamma^{\mu\nu\rho}_i \, p_\mu(\eta) \,\M_{\nu\rho}\otimes 1\right]\triangleright \E_a[\xi]\otimes \E_b[\eta]=
\\
        =&\exp\left[ \,\gamma^{\mu\nu\rho}_i \, p_{\mu}\left(  \xi \cdot \Lambda_i(\eta)  \right) \otimes \M_{\nu\rho}\right] \triangleright \E_a \left[ \xi \cdot \Lambda_i(\eta) \right]\otimes \E_b[\eta]=
\\
=&
\E_a \left[\xi \cdot \Lambda_i(\eta)\right]\otimes \E_b\left[\eta \cdot  \Lambda_i^{-1}\left[\xi \cdot \Lambda_i(\eta)\right) \right]
\,,
    \end{aligned}
\end{equation}
where 
\begin{equation}
(\Lambda_i)^\alpha_{\ \beta}(\xi) =  \exp \left[   -\gamma_i^{\mu\nu\rho}p_\mu(\xi) \, \delta^\alpha_{\ [\nu} \eta_{\rho] \beta} \right]  
\end{equation}
is a 4-dimensional Lorentz matrix whose rapidity/rotation parameters depend on $\xi$ through $p_\mu(\xi)$. Loosely speaking, when acting on a tensor product of plane waves, the factors in the R-matrix perform a sequence of  finite momentum-dependent Lorentz transformations on the plane waves. The end result is two Lorentz-transformed plane waves, with Lorentz transformation parameters that depend (in an increasingly complicated way, depending on the number $n$ in Eq.~\eqref{eq:Rmatgenform}) on the initial momenta $\xi$ and $\eta$.

The full action of the R-matrix on a tensor product of plane waves thus realizes a momentum space diffeomorphism $R : \mathbbm{R}^{3,1} \times \mathbbm{R}^{3,1}  \to \mathbbm{R}^{3,1} \times \mathbbm{R}^{3,1} \,,$ defined by
\begin{equation}
    \label{eq:pwexfinal}\E_b[\eta]\E_a[\xi]=\E_a[R_{(1)}(\xi,\eta)]\E_b[R_{(2)}(\xi,\eta)] \, .
\end{equation}
The inverse map is denoted by $R^{-1}$, in the sense that 
\begin{equation}
    R_{(1)}^{-1}(R_{(1)}(\xi,\eta),R_{(2)}(\xi,\eta))=\xi \qquad  R_{(2)}^{-1}(R_{(1)}(\xi,\eta),R_{(2)}(\xi,\eta))=\eta
\end{equation}
The explicit constructions of the diffeomorphism  $p_\mu(\xi)$ and the expressions for the $R$-matrix in the form of \eqref{eq:Rmatgenform}, from which the expression $R$ can be inferred, are reported in \cref{app:Rmatrices} for all unimodular T-Minkowski models.

\section{Avenues towards quantization and Wick's theorem}
\label{sec:ncqft}

\subsection{Covariant approach to free-field quantization}
\label{sec:covariantquantization}

We begin the discussion concerning quantization of a scalar field theory on T-Minkowski spacetime by reviewing the discussion in the commutative setting. A free commutative real scalar field satisfying the Klein-Gordon equation can be expanded in positive and negative frequencies as 
\begin{equation}
\label{eq:commfieldexpansion}
    \phi(x)=\int \frac{d^3p}{2\omega_{\vec p}}\left[ a(\vec p)e^{-i \, (\omega_p x^0 + \vec p \cdot \vec x)} + a^*(\vec p)e^{i \, (\omega_p x^0 + \vec p \cdot \vec x)}\right] \, ,
\end{equation}
where $\omega_{\vec p}=\sqrt{m^2+\vec{p}^2}$ and $a(p),a^*(p)$ are the Fourier coefficients to be promoted to operators once a suitable quantization scheme is implemented.

The standard procedure employed in the commutative case is that of  canonical quantization, where we assume that the field $\phi(x)$ and its conjugate momentum $\pi(x)$ satisfy the following equal-time commutation relations
\begin{equation}
\label{eq:canonicalquant}
[\phi(t,\vec{x}),\pi(t,\vec{y})]=i\, \delta^{(3)}(\vec{x}-\vec{y}) \, .
\end{equation}
For the case of the free theory, the field and its conjugate momentum satisfy the Hamilton equations
\begin{equation}
    \frac{\partial \phi(t,\vec{x})}{\partial t}=\pi(\vec{x},t) \,, \qquad \frac{\partial \pi(t,\vec{x})}{\partial t}= \Delta \phi(t,\vec{x}) - m^2\phi(t,\vec{x}) \, ,
\end{equation}
which are equivalent to the Klein-Gordon equation for the $\phi(t,x)$, which is thus expanded as in \eqref{eq:commfieldexpansion}.
Once the Fourier coefficients $a(\vec p),a^*(\vec p)$ are promoted to annihilation and creation operators they satisfy the bosonic oscillator algebra 
\begin{equation}
\label{eq:standardhoalgebra}
[a(\vec p),a^\dagger(\vec q)]=2 \, \omega_{\vec p}\,\delta^{(3)}(\vec p-\vec q) \, ,
\end{equation}
in virtue of \eqref{eq:canonicalquant}. This algebra, together with the expression for the field expansion allows to compute the commutator between the scalar field evaluated in two different spacetime points $x$ and $y$, which results in
\begin{equation}
[\phi(x),\phi(y)]=\Delta_{PJ}(x-y) \, , 
\end{equation}
with $\Delta_{PJ}(x-y)$ being the Pauli-Jordan function, whose expression reads
\begin{equation}
\label{eq:commpj}
    \Delta_{PJ}(x-y)=\int \frac{d^3p}{2\omega_{\vec p}} \,\left[ e^{-i\, p_\mu  (x-y)^\mu}-e^{i\, p_\mu  (x-y)^\mu} \right]\,, \qquad \text{where~~} p_0 = \omega_{\vec p}\,.
\end{equation}
The development of the free theory presented above then allows to calculate $N$-point correlation functions, completely determined from the $2$-point function, for both the free 
and interaction cases; the latter by means of Hamiltonian perturbation theory.

A noncommutative generalization of the quantization program outlined above is bound to face significant obstructions. Indeed, the definition of the conjugate momentum $\pi(t,\vec{x})$ requires the development of Hamiltonian dynamics, which in turn requires the definition of an equal-time hypersurface (or of a more general spacelike hypersurface of simultaneity).  In the noncommutative setting where, in most models, the time variable is noncommutative, a hypersurface of constant time might not exist, in the sense that there is no spacelike region that can be localized arbitrarily well in time.

\subsection{Covariant approach to free-field quantization}

A different quantization program that lends itself naturally to a noncommutative generalization is
as described in \cite{schweber2013introduction}. This approach rests upon three basic assumptions in the commutative case:
\begin{enumerate}
    \item The scalar field $\phi(t,\vec{x})$ satisfies the Klein-Gordon equation, $(\square+m^2)\phi(t,\vec{x})=0$.
    \item The commutator between fields evaluated at different spacetime points is a c-number,  $[\phi(x),\phi(y)]=F(x,y)$.
    \item The function $F(x,y)$ is Poincaré invariant, in the sense that $F(\Lambda \, x+a, \Lambda \, y+a)=F(x,y)$.
\end{enumerate}
In chapter 7c of \cite{schweber2013introduction} it is shown that these requirements are sufficient to prove that $F(x,y)$ is proportional to the Pauli Jordan function $\Delta_{PJ}(x-y)$. Once the commutator between fields in different spacetime points is known, the scalar field can be expanded in modes to be quantized according to the commutation relations derived by computing $[\phi(x),\phi(y)]$ explicitly, leading to \eqref{eq:standardhoalgebra}. Then one is equipped with all the tools to compute physical predictions for both free and interacting QFTs.

% \subsection{Covariant approach in the noncommutative setting}

The quantization prescription described in \cite{schweber2013introduction} for the commutative theory can be generalized to the noncommutative framework in the following way.  When quantizing the momentum-dependent Fourier coefficients, the scalar field becomes an element of the tensor product of the noncommutative algebra of functions and of the operators on the Fock space, namely, $\phi(\x)\in\mathcal{O}(\mathcal{H})\otimes C_\ell[\mathbbm{R}^{3,1}]^{\otimes_\ell N}$\footnote{We are implicitly assuming that the creation and annihilation operators commute with the space-time coordinates. This hypothesis has been relaxed in \cite{Fiore:2007vg} in the study of non-commutative QFT on Moyal space, leading to a different realization of commutation relations between creation and annihilation operators, without however changing the physical content of the theory.}. Then, the list of assumptions of chapter 7c of \cite{schweber2013introduction} generalizes to 
\begin{enumerate}
    \item The scalar field $\phi(\x)$ satisfies the noncommutative version of the Klein-Gordon equation  \eqref{eq:NCKG}, $\square \triangleright \phi(\x)+m^2\phi(\x)=0 \, .$
    \item The commutator $[\phi(\x_a),\phi(\x_b)]$ is an element of  $1\otimes C_\ell[\mathbbm{R}^{3,1}]^{\otimes_\ell N}\simeq C_\ell[\mathbbm{R}^{3,1}]^{\otimes_\ell N} $ and equal to a generic non-commutative function $F(\x_a,\x_b) \in C_\ell[\mathbbm{R}^{3,1}]^{\otimes_\ell N}$.
    \item The function $F(\x_a,\x_b)$ is invariant under T-Minkowski transformations, namely $F(\x'_a,\x'_b)=F(\x_a,\x_b)$ where $\x_a^{'\mu}=\Lambda^\mu{}_\nu\otimes \x^\nu + a^\mu\otimes 1$. 
\end{enumerate}

Tracing the steps of the proof in  \cite{schweber2013introduction}, we now show that the three axioms for covariant quantization, adapted to the noncommutative case, imply that $F(\x_a,\x_b)$ is equivalent to the commutative Pauli-Jordan function.

With the formalism developed in section \ref{matprel}, the first of these requirements  allows us to expand a real scalar field as \footnote{The expression \eqref{eq:phiexpcov} is valid for any T-Minkowski real scalar field satisfying the noncommutative Klein-Gordon equation. This includes models such as lightlike $\kappa$-Minkowski \cite{Lizzi:2021rlb,DiLuca:2022idu,Fabiano:2023xke} where also "new type" plane waves, defined on previously-ignored subregions of momentum space, appear in the expansion of the scalar field. Even in such cases, the scalar field expansion can be formally written as in \eqref{eq:phiexpcov}.}
\begin{equation}
\label{eq:phiexpcov}
    \phi(\x_a)=\int d^4\xi \,\delta(\xi^2-m^2)\Tilde{\phi}(\xi)\E_a[\xi] \, .
\end{equation}
The second requirement simply states that
\begin{equation}
\label{eq:phixphiy}
    [\phi(\x_a),\phi(\x_b)]=F(\x_a,\x_b) \, ,
\end{equation}
where $F(\x_a,\x_b)$ is a generic function of multiple coordinates that we can write as 
\begin{equation}
\label{eq:Fgeneral}
    F(\x_a,\x_b)=\int d^4\xi d^4\eta\,  \Tilde{F}(\xi,\eta)\E_a[\xi]\E_b[\eta] \, .
\end{equation}
The T-Poincaré invariance of $F(\x_a,\x_b)$ implies that it can be written in one of the forms displayed in \eqref{eq:2pointdistr}. For definiteness, we choose the first of these option, expanding $F$ in terms of $\E_a[\xi]\E_b^*[\xi]$, thus requiring that $\Tilde{F}(\xi,\eta)=\delta(\eta-S(\xi))\Tilde{F}(\xi)$. Therefore, the function $F(\x_a,\x_b)$ can then be written as
\begin{equation}
    F(\x_a-\x_b)=\int d^4\xi \, \Tilde{F}(\xi)\E_a[\xi]\E_b^*[\xi]
\end{equation} 
Using the expression for the scalar field \eqref{eq:phiexpcov}, we notice that the left hand side of \eqref{eq:phixphiy} is solely dependent on on-shell plane waves. As a consequence, $F(\x_a-\x_b)$ is non-zero only when the plane waves on which it depends are also on-shell, so we can write $\Tilde{F}(\xi)=\Tilde{f}(\xi)\delta(\xi^2-m^2)$, and thus 
\begin{equation}
\label{eq:fspecialized}
   F(\x_a-\x_b)=\int d^4\xi \, \Tilde{f}(\xi)\E_a[\xi]\E_b^*[\xi]\delta(\xi^2-m^2)=\int d^4\xi \, \Tilde{f}(\xi) e^{i\xi(\x_a-\x_b)}\delta(\xi^2-m^2) \, .
\end{equation}

Another consequence of condition 3 is that the function $F(\x_a-\x_b)$ must be 
Lorentz invariant, which explicitly reads
\begin{equation}
    \begin{aligned}        F(\x_a^\prime-\x_ b^\prime)=&\int d^4\xi e^{i\xi_\mu(\x_a^\prime-\x_b^\prime)^\mu}\delta(\xi^2-m^2)\Tilde{f}(\xi)=\int d^4\xi e^{i\xi_\mu\hat{\Lambda}^\mu{}_\nu(\x_a^{\nu}-\x_b^{\nu})}\delta(\xi^2-m^2)\Tilde{f}(\xi)=\\
    =&\int d^4\xi e^{i\xi_\mu(\x_a-\x_b)^\mu}\delta(\xi^2-m^2)\tilde{f}(\xi_\nu (\hat{\Lambda}^{-1})^\nu{}_\mu)=\int d^4\xi \, e^{i\xi_\mu(\x_a-\x_b)^\mu}\delta(\xi^2-m^2)\Tilde{f}(\xi) \, ,
    \end{aligned}
\end{equation}
and implies that $\tilde{f}(\xi)$ must be a Lorentz invariant function. Moreover, the on-shell condition implies that $\xi_\mu$
must be a time-like four vector, so $\Tilde{f}(\p)$ can only be a function of $\xi^2$ and $\epsilon(\xi_0)=\theta(\xi_0)-\theta(-\xi_0)$, which are the only two Lorentz invariants one can form with a time-like four vector. Therefore we posit
\begin{equation}
\Tilde{f}(\xi)=\Tilde{f}_1(\xi^2)+\epsilon(\xi_0)\Tilde{f}_2(\xi^2) \, . 
\end{equation}
When plugging this in \eqref{eq:fspecialized}, the functions $\Tilde{f}_1$ and $\Tilde{f}_2$ both turn into functions of $m^2$, because of the delta function imposing the on-shell condition, leaving us with
\begin{equation}
\label{eq:Fformadef}
    F(\x_a-\x_b)=\int d^4\xi \, \E_a[\xi]\E_b^*[\xi]\delta(\xi^2-m^2)\left[\tilde{f}_1(m^2)+\epsilon(\xi_0)\tilde{f}_2(m^2)\right] \, .
\end{equation}
From the definition \eqref{eq:phixphiy} and using the fact that $\phi(\x_a)$ is a real scalar field, it is easy to see that $F(\x_b-\x_a)=-F(\x_a-\x_b)$ and that $F^*(\x_a-\x_b)=-F(\x_a-\x_b)$. Imposing these conditions on \eqref{eq:Fformadef}, we conclude that $f_1=0$ and that $f_2=\overline{f_2}$, so that $f_2$ must be real.

We are finally left with the expression
\begin{equation}
    F(\x_a-\x_b)=f_2\int d^4\xi \, e^{i\xi_\mu(\x_a^\mu-\x_b^\mu)}\delta(\xi^2-m^2)\epsilon(\xi_0) \, .
\end{equation}
which, up to constant factor $f_2$, coincides with the commutative Pauli-Jordan function \eqref{eq:commpj}, since coordinate differences are commutative. The commutative limit then requires us to set $f_2=-1$, and we can write
\begin{equation}
\label{eq:ncpj}
    F(\x_a-\x_b)=\Delta_{PJ}(\x_a-\x_b) ,
\end{equation}
The commutation relation between noncommutative quantum scalar fields evaluated in different points, for any T-Minkowski model, is thus given by
\begin{equation}
    \label{eq:phiphipj}[\phi(\x_a),\phi(\x_b)]=\Delta_{PJ}(\x_a-\x_b) \, .
\end{equation}
When the expression for the scalar field $\eqref{eq:phiexpcov}$ is further refined in terms of creation and annihilation modes, the explicit computation of \eqref{eq:phiphipj} allows us to compute the commutation relations between these operators, which, as we will see in the next section, will turn out to be deformed due to the non-trivial commutation relations of plane waves evaluated in different points.

\subsection{Noncommutative N-point functions and Wick's theorem}
\label{sec:wicktheorem}
While the result from the previous section is valid for any T-Minkowski model, we now specialize to unimodular cases, characterized by $\left|\det \frac{\partial S(\xi)}{\partial \xi }\right|=1$, and corresponding to cases 10 to 16 and 19 to 21 in the classification of \cite{tolstoy2007twistedquantumdeformationslorentz}. As we will see, this assumption allows us to conveniently further expand the scalar field expression $\eqref{eq:phiexpcov}$ solely in terms of noncommutative generalizations of positive and negative frequency modes and to prove a noncommutative version of the Wick theorem. As already shown in \cite{DiLuca:2022idu,Fabiano:2023xke} for lightlike 
$\kappa$-Minkowski, the expression for the scalar field in the non-unimodular cases is not simply expandable in terms of positive and negative frequency modes but requires the introduction of new-type plane waves, which in turn leads to the introduction of new quantum operators in addition to the usual creation and annihilation operators. The complete characterization of the N-point functions for these non-unimodular models will be developed in a future dedicated study.

Focusing now on the unimodular case, we can enforce the on-shell constraint in \eqref{eq:phiexpcov} and rewrite the expression for the scalar field as 
\begin{equation}
\label{eq:filedexpcalculation}
    \begin{aligned}
\phi(\x_a)=&\int_0^{\infty}d\xi_0\int d^3\xi \, \delta(\xi^2-m^2) \Tilde{\phi}(\xi)\E_a[\xi]+\int_{-\infty}^0 d\xi_0\int d^3\xi \, \delta(\xi^2-m^2) \Tilde{\phi}(\xi)\E_a[\xi]=\\
    =&\int_0^{\infty}d\xi_0\int d^3\xi \, \delta(\xi^2-m^2)\left[ \Tilde{\phi}(\xi)\E_a[\xi]+ \abs{\det \left( \frac{\partial S(\xi)}{\partial \xi}\right)} \Tilde{\phi}(S(\xi))\E_a^*[\xi]\right]=\\
    =&\int \frac{d^3\xi}{2\omega_{\vec \xi}}\left[\Tilde{\phi}(\xi)\e_a[\vec \xi]+\Tilde{\phi}(S(\xi))\e_a^*[\vec \xi]\right] \, , 
    \end{aligned}
\end{equation}
where in the second equality we have performed the change of variables $\xi\rightarrow S(\xi)$ and have used the fact that $\abs{\det \partial_\xi S(\xi)}=1$ since the model is unimodular. Moreover, the second equality also implies that the change of variables $\xi\rightarrow S(\xi)$ maps the region $\xi_0>0$ into the $\xi_0<0$ region and that the spatial component of the four-momentum $\vec \xi$ spanning all $\mathbb{R}^3$ is mapped into $S(\vec \xi)$ spanning all $\mathbb{R}^3$ as well. This is indeed not trivial since $S(\xi)_\mu=-\xi_\nu(\chi^{-1})^\nu{}_\mu(\xi)$, with $(\chi^{-1})^\nu{}_\mu(\xi)$ being an $SO(3,1)$ matrix depending on $\xi$ itself. However, it can be checked that these properties hold for all the unimodular cases: in \cref{app:Rmatrices} we show a method to explicitly verify this. In the third equality, we have set
\begin{equation}
    \omega_{\vec \xi}=\sqrt{m^2+\vec{\xi}^2} \,.
\end{equation}
 
The reality condition $\phi^*(\x_a)=\phi(\x_a)$ requires that $\overline{\Tilde{\phi}(\xi)}=\Tilde{\phi}[S(\xi)]$. By denoting $\Tilde{\phi}[S(\xi)]=a(\xi)$, and promoting $a(\vec \xi)$ and $a^*(\vec \xi)$ to quantum operators, the expression for the real scalar field becomes
\begin{equation}
\label{eq:fieldexpansion}
    \phi(\x_a)=\int \frac{d^3\xi}{2\omega_\xi} \left[ a(\vec \xi)\e_a^*[\vec \xi]+a^\dagger(\vec \xi)\e_a[\vec \xi]\right] \, .
\end{equation}
and the commutation relations between these operators can be found by computing \eqref{eq:phiphipj} explicitly, whose right-hand side involves the non-commutative generalization of the Pauli Jordan function which in the unimodular case can be cast in the form
\begin{equation}
\label{eq:ncpj}
    \Delta_{PJ}(\x_a-\x_b)=\int \frac{d^3\xi}{2\omega_{\vec \xi}}\left(\e_b[\vec \xi]\e_a^*[\vec \xi]-\e_a[\vec \xi]\e_b^*[\vec \xi]\right) \, .
\end{equation}
The left-hand side of \eqref{eq:phiphipj} requires not only to take into account the non-commutativity of the operators $a(\vec \xi)$ and $a^\dagger(\vec \xi)$ but also that of plane waves evaluated in different points, which is the main difference with respect to the commutative case, and generally leads to a deformation of the harmonic oscillator algebra \cite{Fiore:2007vg,Fabiano:2023xke}.  
In \cref{app:oscillatoralgebra} we present details of the explicit computation of \eqref{eq:phiphipj}, handling noncommutativity of plane waves as in \eqref{eq:pwexfinal}, using the expressions for the $R$-matrices in \cref{app:Rmatrices}. For all unimodular T-Minkowski models with the cocycle $\theta^{\mu\nu}$ set to 0, the deformed harmonic oscillator algebra can be written in the following form 
\begin{equation}
\label{eq:deformedoscillatoralgebra}
    \begin{aligned}
    &a(\xi)a(\eta)-a(R_{(1)}^{-1}(\eta,\xi))a(R_{(2)}^{-1}(\eta,\xi))=0\\
        &a^\dagger(\xi)a^\dagger(\eta)-a^\dagger(R_{(2)}^{-1}(\xi,\eta))a^\dagger(R_{(1)}^{-1}(\xi,\eta))=0\\
        &a(\xi)a^\dagger(\eta)-a^\dagger(R_{(2)}^{-1}(S(\xi),\eta))a(S(R_{(1)}^{-1}(S(\xi),\eta)))=2\omega_\xi\delta(\vec{\xi}-\vec{\eta}) \, ,
    \end{aligned}
\end{equation}
In \eqref{eq:deformedoscillatoralgebra}, the arguments $(\xi,\eta)$ of the components of the R-matrix function are to be understood as $(\omega_{\vec \xi},\vec \xi,\omega_{\vec \eta},\vec \eta)$, since $R$ is applied on 4-vectors identifying on-shell plane waves.

By defining the vacuum state $\ket{0}$ as the T-Poincaré invariant state that is annihilated by $a(\xi)$, we have all the information needed to compute the $N$-point functions of the free theory, which we define as the vacuum expectation values of products of $N$ fields evaluated in different points. This is none other than the noncommutative generalization of the definition of Wightmann functions, which has also been used in \cite{Fiore:2007vg} to investigate QFT on the Moyal spacetime. In formulas 
\begin{equation}
\label{eq:wightmann}
    W(\x_1,\dots,\x_n)\coloneqq \bra{0}\phi(\x_1) \dots \phi(\x_n)\ket{0} \, .
\end{equation} 
For convenience, we can distinguish positive and negative frequency parts of the scalar field decomposition, denoting them by
\begin{equation}
    \phi_+(\x_a)=\int \frac{d^3\xi}{2\omega_{\vec \p}} a(\vec \p)\e_a^*[\vec \p] \, , \qquad \phi_-(\x_a)=(\phi_+(\x_a))^\dagger = \int \frac{d^3\xi}{2\omega_{\vec \p}} a^\dagger(\vec \p)\e_a[\vec \p] \, , 
\end{equation}
respectively, and it is implied that the $\dagger$ operation takes the hermitian conjugate of operators on $\mathcal{O}(\mathcal{H})$ and the involution on elements of $C_\ell[\mathbbm{R}^{3,1}]^{\otimes_\ell N}$. From these definitions we immediately deduce that $\phi_+(x_a)\ket{0}=0$ and $\bra{0}\phi_-(x_a)=0$, so that the two point Wightmann function $W(\x_a,\x_b)=\bra{0}\phi(\x_a)\phi(\x_b)\ket{0}$ simplifies to $\bra{0}\phi_+(\x_a)\phi_-(\x_b)\ket{0}$, which we can compute starting from
\eqref{eq:deformedoscillatoralgebra}. Explicitly, we have
\begin{equation}
\label{eq:wightmann}
    \begin{aligned}       W(\x_a,\x_b)=\bra{0}\phi_+(\x_a)\phi_-(\x_b)\ket{0}=&\int \frac{d^3\xi}{2\omega_\xi}\frac{d^3\eta}{2\omega_\eta}\bra{0}a(\xi)a^\dagger(\eta)\ket{0}\e_a^*[\xi]\e_b[\vec\eta]=\\
        =&\int \frac{d^3\xi}{2\omega_\xi}\frac{d^3\eta}{2\omega_\eta}2\omega_\xi\delta(\vec{\xi}-\vec{\eta})\e_a^*[\xi]\e_b[\vec\eta]=\\
        =&\int \frac{d^3\xi}{2\omega_\xi}\e_a^*[\vec\xi]\e_b[\vec\xi] \, . 
    \end{aligned}
\end{equation}
The function obtained is translation invariant since its defining integral depends on the product $e^*_a[\xi]e_b[\xi]$. Therefore, in the following, we will write $W(\x_a,\x_b)=W(\x_a-\x_b)$.
By recasting the last integral of \eqref{eq:wightmann} in a covariant form and performing the change of variables $\xi\rightarrow S(\xi)$,
the result can also be written as 
\begin{equation}
\label{eq:wightmann2}
\begin{aligned}  W(\x_a-\x_b)=&\int_0^{\infty}d\xi_0\int d^3\xi \, \delta(\xi^2-m^2)\E_a^*[\xi]\E_b[\xi]=\int_{-\infty}^0 d\xi_0\int d^3\xi \,\delta(\xi^2-m^2)\E_a[\xi]\E_b^*[\xi]=\\
    =&\int_{-\infty}^0 d\xi_0\int d^3\xi \,\delta(\xi^2-m^2)e^{i\xi(\x_a-\x_b)}=\int_{0}^\infty d\xi_0\int d^3\xi \,\delta(\xi^2-m^2)e^{i\xi(\x_b-\x_a)}=\\
    =& \int \frac{d^3\xi}{2\omega_{\vec \xi}}\e_b[\vec \xi]\e_a^*[\vec \xi]  \, .
\end{aligned}
\end{equation}
Focusing on the expression for the 2-point Wightmann function in the second to last equality of \eqref{eq:wightmann2}, it is easy to check its T-Poincaré invariance.
Indeed
\begin{equation}
\begin{aligned}
    W(\x'_a-\x'_b)=&\int_{0}^\infty d\xi_0\int d^3\xi \,\delta(\xi^2-m^2)e^{i\xi(\x'_b-\x'_a)}=\\
    =&\int_{0}^\infty d\xi_0\int d^3\xi \,\delta(\xi^2-m^2)e^{i\xi_\mu \,\La^\mu{}_\nu \otimes (\x_b-\x_a)^\nu} 
    \end{aligned}
\end{equation}
By performing the change of variables $\xi_\nu'=\xi_\mu\La^\mu{}_\nu$, which simply amounts to the standard linear Lorentz transformation for momenta, it is easy to see that the measure, the integration region and the on-shell condition are left invariant, resulting in
\begin{equation}
    W(\x'_a-\x'_b)=  \1 \otimes W(\x_a-\x_b) \, .
\end{equation}
Using the last expression of $W(\x_a-\x_b)$ in \eqref{eq:wightmann2} one can also show that
\begin{equation}
    W(\x_a-\x_b)-W(\x_b-\x_a)=\Delta_{PJ}(\x_a-\x_b) \, ,
\end{equation}
which is the standard relation between the Pauli-Jordan function and the Wightmann functions one also encounters in the commutative case.

The explicit computation of higher order $N$-point functions can become quite a cumbersome task since it involves both the commutation relations between creation and annihilation operators and the commutation relations between plane waves living in different copies of $C_\ell[\mathbbm{R}^{3,1}]$.
We will now prove a noncommutative version of Wick's theorem which simplifies this task and predicts that all $N$-point functions can be written in terms of products of two-point Wightmann functions. In other words, since the two-point Wightmann function is a commutative function, equal to the one from the commutative case and since coordinate differences all commute among themselves, then all $N$-point functions of the free theory are going to be equal to the ones of the commutative theory.

The generic $N$-point function, as defined in \eqref{eq:wightmann}, can be written as a sum of contributions of the type
\begin{equation}
\label{eq:wick0}
    \bra{0}\phi_{\alpha_1}(\x_1)\dots\phi_{\alpha_N}(\x_N)\ket{0} \, ,
\end{equation}
where $\alpha_i\in\{+,-\}$.
The general strategy to compute these terms is similar to the one adopted in the commutative case: the objective is to commute annihilation operators to the far right side of the expression, so that they annihilate the vacuum, producing two-point Wightmann functions along the way when an annihilation operator commutes with a creation operator. The additional level of complexity in this case, other than the deformation of the oscillator algebra written in \eqref{eq:deformedoscillatoralgebra}, is the fact that plane waves are non-commutative objects.

To tackle this issue, we first observe that as a direct consequence of the computations developed in \Cref{app:oscillatoralgebra} to obtain the deformed oscillator algebra \eqref{eq:deformedoscillatoralgebra}, we have 
\begin{equation}
\label{eq:phi+phi-}
    [\phi_+(\x_a),\phi_-(\x_b)]=W(\x_a-\x_b) \, .
\end{equation}
Also, using the first result proved in \Cref{subsec:theorems}, for $h(\x_b-\x_c)=W(\x_b-\x_c)$, we deduce that the Wightmann function produced by commuting $\phi_+(\x_b)$ with $\phi_-(\x_c)$ can always be factorized from the expression \eqref{eq:wick0}, \textit{i.e.}, the Wightmann function commutes with all plane waves (its commutation with annihilation and creation operators is trivial, given the assumptions of our model). To show this, consider 
\begin{equation}
\begin{aligned}
    \E_a[\eta]W(\x_b-\x  _c)&=\int_0^\infty d\xi_0\int d^3\xi \, E_a[\eta] E_c[\p]E_b^*[\p]\delta(\xi^2-m^2)=\\
    &=\int_0^\infty d\xi_0\int d^3\xi \E_a[\eta]\, e^{-i \p_\mu(\x_b-\x_c)^\mu}\delta(\xi^2-m^2)=\\
    &=\int_0^\infty d\xi_0\int d^3\xi\, e^{-i\xi_\mu\chi^\mu{}_\nu(\eta) (\x_b-\x_c)^\nu}\E_a[\eta]\delta(\xi^2-m^2)=\\
    &=\int_0^\infty d\xi_0\int d^3\xi\, e^{-i\xi(\x_b-\x_c)}\E_a[\eta]\delta(\xi^2-m^2)= W(\x_b-\x_c) \E_a[\eta] \, . 
\end{aligned}
\end{equation}
In the derivation above, we have used the fact that $\chi^\mu{}_\nu(\eta)$ is an $SO(3,1)$ element so that the integration measure and the on-shell condition remain invariant under the change of variables $\xi_\mu\rightarrow \chi^\mu{}_\nu(\eta)\xi_\nu$.

The two conditions
\begin{equation}
\label{eq:intermediateresults}
    [\phi_+(\x_a),\phi_-(\x_b)]=W(\x_a-\x_b) \, , \qquad
    [\E_a[\eta],W(\x_b-\x_c)]=0 \,,
\end{equation}
are sufficient to prove our result for the Wick theorem. Consider a contribution of the form \eqref{eq:wick0} with an equal number of ``+'' and ``-'' fields. Adopting the strategy outlined above, whenever a $\phi_+(\x_i)\phi_-(\x_j)$ term occurs, one can use relations \eqref{eq:intermediateresults} to write 
\begin{equation}
\begin{aligned}
    &\bra{0}\phi_{\alpha_1}(\x_1)\dots \phi_{\alpha_{i-1}}(\x_{i-1})  \phi_+(\x_i)\phi_-(\x_{i+1}) \phi_{\alpha_{i+1}}(\x_{i+1})\dots \phi_{\alpha_N}(\x_N)\ket{0}= \\
    &=W(\x_i,\x_{i+1})\bra{0}\phi_{\alpha_1}(\x_1)\dots \phi_{\alpha_{i-1}}(\x_{i-1})\phi_{\alpha_{i+1}}(\x_{i+1})\dots \phi_{\alpha_N}(\x_N)\ket{0}
    \\
    &~~+\bra{0}\phi_{\alpha_1}(\x_1)\dots \phi_-(\x_{i+1})\phi_+(\x_{i})\dots \phi_{\alpha_N}(\x_N)\ket{0} \,. 
\end{aligned}
\end{equation}
The vacuum expectation value in the first term of this expression now contains $N-2$ fields, since two of them have been contracted to produce $W(\x_i-\x_{i+1})$, while the second term still contains $N$ fields but a $\phi_+$ term has migrated to the right. The process just discussed above can then be reiterated, producing multiple terms until one is left with only non-zero terms made up exclusively of products of two-point functions.
It is also straightforward to apply this reasoning to cases in which there is an imbalance between the $+$  and $-$ terms, or to cases in which there are an odd number of fields, to eventually find out that they are all zero.

Alas, we obtain the expression for the $N$-point Wightman functions
\begin{equation}
\label{eq:npointfunction}W(\x_1,\dots,\x_N)=\sum_{\substack{\text{pairing} \\ \text{of fields}}}\prod_{\text{pairs}}W(\x_i-\x_j) \, , 
\end{equation}
which is identical to the commutative one. In light of \eqref{eq:npointfunction}, the T-Poincaré invariance of the $N$-point functions is guaranteed from the T-Poincaré invariance of the Wightmann function. In commutative QFT, a direct consequence of \eqref{eq:npointfunction} is a Wick theorem for  $N$-point Green functions, defined as vacuum expectation values of time ordered products of $N$-fields. As we will see in \cref{sec:interactions}, a generalization of this definition in the non-commutative case will be available only for a subset of unimodular T-Minkowski models.

We will now proceed to exhibit the properties of some specific $T$-Minkowski models with which can be used to explicitly verify the result we have shown in this section.

\subsection{Specific examples}

\subsubsection{$\theta$-Minkowski} 
\label{sec:theta}
The triviality of $N$-point functions for a covariant QFT on $\theta$-Minkowski has been proved in \cite{oeckl2000untwisting,Fiore:2007vg}, using the star product formalism and more recently in~\cite{Bogdanovic:2024jnf}, using the Batalin–Vilkovisky  and  $L_\infty$ formalisms. We review this result within our framework,
with the intention of also showing how to deal with noncommutativity of pure central extension type, covering cases 19, 20, 21 listed in \cref{app:Rmatrices}. 
The coordinate commutators are expressed by
\begin{equation}
    \label{eq:thetanoncomm}[\x^\mu,\x^\nu]=i\theta^{\mu\nu} \, ,
\end{equation}
where $\theta^{\mu\nu}$ is a real-valued, constant, anti-symmetric matrix, understood as multiplied by the identity element of the algebra, in the commutators. The braided tensor product algebra and the differential calculus are defined by the relations \cite{Fiore:2007vg,Mercati:2023apu}
\begin{equation}
\label{eq:thetabraidedalgebra}
[\x_a^\mu,\x_b^\nu]=i\theta^{\mu\nu} \, , \qquad [\x_a^\mu,\x_b^\nu-\x_c^\nu]=[\x_a^\mu,d\x^\nu]=0 \, ,
\end{equation}
which imply that all the differential calculus structures are undeformed.  
One can define noncommutative plane waves using the Weyl ordering
\begin{equation}
    \E_a[\p]=e^{i\p_\mu \x^\mu} \, ,\qquad \E_a[S(\xi)]=e^{-i\xi_\mu\x^\mu}\,.
\end{equation}
with coordinates $\xi_\mu$ transforming linearly under Lorentz transformations \footnote{Other orderings simply amount to multiplying the plane wave by a momentum-dependent phase factor, due to the noncommutativity structure \eqref{eq:thetanoncomm}. As can be seen from the remainder of the discussion, this would not affect the expression of N-point functions.}
The R-matrix for this model is
\begin{equation}
\mathcal{R}=e^{i\theta^{\mu\nu}\X_\mu\otimes \X_\nu} \, ,
\end{equation}
with $\X_\mu\triangleright \E[\xi]=\xi_\mu\E[\xi]$, and regulates the exchange of plane waves of different points, as follows
\begin{equation}
\label{eq:thetaexchange}
    \E_b[\eta]\E_a[\xi]=\mu\circ R\triangleright \E_a[\xi]\otimes \E_b[\eta]=e^{i\p_\mu\theta^{\mu\nu}\eta_\nu}\E_a[\p]\E_b[\eta] \, .
\end{equation}
The scalar field of the theory can be expanded as in \eqref{eq:fieldexpansion} and through \eqref{eq:phiphipj} one finds the deformed harmonic oscillator algebra, as in \cite{Fiore:2007vg}:
\begin{equation}
\label{eq:thetadho}
\begin{aligned}
&a(\vec \p)a(\vec \eta)=e^{i\p_\mu\theta^{\mu\nu}\eta_\nu}a(\vec \eta)a(\vec \p) \\
&a^\dagger(\vec \p)a^\dagger(\vec \eta)=e^{i\p_\mu\theta^{\mu\nu}\eta_\nu}a^\dagger(\vec \eta)a^\dagger(\vec \p) \\
&a(\vec \p)a^\dagger(\vec \eta)-e^{-i\p_\mu\theta^{\mu\nu}\eta_\nu}a^\dagger(\vec \eta)a(\vec \p)=2\omega_\p\delta(\vec \p-\vec \eta) \, ,
\end{aligned}
\end{equation}
where it is implied that $\xi_0=\omega_{\vec \xi}=\sqrt{\vec \xi^2+m^2}$ and $\eta_0=\omega_{\vec\eta}=\sqrt{\vec\eta^2+m^2}$. 
Making use of \eqref{eq:thetaexchange} and \eqref{eq:thetadho}, we can calculate [$\phi_+(\x_a),\phi_-(\x_b)$]:
\begin{equation}
\label{eq:phiaphibtheta}
\begin{aligned}
    \phi_+(\x_a)\phi_-(\x_b)=&\int \frac{d^3\xi}{2\omega_\xi}\frac{d^3\eta}{2\omega_\eta} a(\xi)a^\dagger(\eta)E_a^\dagger[\xi]E_b[\eta]=\\
    =&\int \frac{d^3\xi}{2\omega_\xi}\frac{d^3\eta}{2\omega_\eta}\left( e^{-i\xi_\mu\theta^{\mu\nu}\eta_\nu}a^\dagger(\eta)a(\xi)+2\omega_\xi\delta(\xi-\eta)\right)E_a^\dagger[\xi]E_b[\eta] = \\
    =&\int \frac{d^3\xi}{2\omega_\xi}\frac{d^3\eta}{2\omega_\eta}a^\dagger(\eta)a(\xi)E_b[\eta]E_a^\dagger[\xi]+W(\x_a-\x_b)=\phi_-(\x_b)\phi_+(\x_a) + W(\x_a-\x_b) \,. 
\end{aligned}
\end{equation}
Furthermore, given the fact that coordinate differences are central, Eq. \eqref{eq:thetabraidedalgebra}, we immediately deduce that 
\begin{equation}
\label{eq:phiWtheta}
    [\phi_{\pm}(\x_a),W(\x_b-\x_c)]=0 \, .
\end{equation}
As argued in \Cref{sec:wicktheorem}, relations \eqref{eq:phiaphibtheta} and \eqref{eq:phiWtheta} are sufficient to show that the undeformed Wick theorem holds.

\subsubsection{$\zeta$-Minkowski ($\varrho$- and $\lambda$-Minkowski)} 

These three models~\cite{Lukierski:2005fc,Lizzi:2022hcq,Fabiano:2023uhg}, collectively indicated as $\zeta$-Minkowski in~\cite{Mercati:2023apu} (initially called ``Lie-Algebraic'' in~\cite{Lukierski:2005fc}), have different names according to the sign of the Lorentzian square of the vector $\zeta^\mu$: timelike is $\varrho$,  spacelike is $\lambda$ and lightlike has no letter assigned yet. They correspond to cases 13, 14, 15, respectively. 
The commutation relations are
\begin{equation}
\label{eq:rhocommrel}
   \begin{aligned}
& [\x^0,\x^1]=i \x^2 \zeta _0{}^{(\varrho )}\,,& &[\x^0,\x^2]=-i \x^1 \zeta _0{}^{(\varrho )}\,,& &[\x^1,\x^2]=0\,,
 \\ 
 & [\x^0,\x^3]=0\,,& &[\x^1,\x^3]=i \x^2 \zeta _3{}^{(\varrho )}\,,& &[\x^2,\x^3]=-i \x^1 \zeta _3{}^{(\varrho )}\,,
   \end{aligned}
\end{equation}
where $\zeta_\mu^{(\varrho)}=\delta_\mu^0$ for the timelike version, $\zeta_\mu^{(\varrho)}=\delta_\mu^0+\delta_\mu^3$ for the lightlike version, and  $\zeta_\mu^{(\varrho)}=\delta_\mu^3$ for the spacelike version.

These models are much less studied than $\theta$-Minkowski. For $\lambda$-Minkowski, we are only aware of~\cite{DimitrijevicCiric:2018blz}, where a non-covariant quantization scheme is applied to a scalar QFT with quartic interaction, demonstrating IR/UV mixing properties (\textit{i.e.} new IR divergencies that are absent in the commutative theory) in the calculation of the one-loop correction of the 2-point function. The non-planar diagrams violate the conservation of momentum and Lorentz invariance.
In the case of $\varrho$-Minkowski, the recent~\cite{Hersent:2023lqm} studied non-covariant complex scalar QFT with quartic interaction, and again IR/UV mixing terms that violate Poincaré covariance are found. These are always present in the one-loop corrections to the 4-point function and also, depending on the form of the interaction term (a complex noncommutative theory admits two), in the one-loop 2-point function.

These result parallel those obtained in the past when IR/UV mixing was first discovered in $\theta$-Minkowski~\cite{Minwalla:1999px}, and are a consequence of a quantization scheme that is incompatible with the deformed Poincaré invariance of the relevant noncommutative spacetime. With the covariant scheme presented in this paper, we are able to show that, both at the free and at the perturbative interactive level, these theories are free of IR/UV mixing and their N-point functions are all undeformed, \textit{i.e.} identical to their commutative counterparts.

Here, we will briefly illustrate the calculations for the free theory, for any choice of the vector $\zeta^\mu$. In the following section we will complete the analysis with an argument showing the triviality (at the perturbative level) of interacting theories.

We choose time-to-the-right ordering for plane waves
\begin{equation}
    \E[p]=e^{ip_i\x^i}e^{ip_0\x^0} \, ,
\end{equation}
such that for all three cases we have $\xi_\mu(p)=p_\mu$ (see \cref{app:Rmatrices} for details) and 
\begin{equation}
    S(\xi)_\mu=\left(-\xi_0,-\xi_1\cos(\tilde{\xi})-\xi_2\sin(\tilde{\xi}),-\xi_2\cos(\tilde{\xi})+\xi_1\sin(\tilde{\xi}),-\xi_3\right) \, ,
\end{equation} 
where $\tilde{\xi}=\zeta_\mu^{(\rho)}\eta^{\mu\nu}\xi_\nu$ and is equal to $-\xi_0,-\xi_0+\xi_3$ or $\xi_3$ wether we are in the timelike, lightlike or spacelike case, respectively. 
Then the R-matrix can be written for all three cases in a compact way in terms of the $\X_\mu$ translation generators as 
\begin{equation}
    \mathcal{R}=e^{\zeta_\mu^{(\rho)}\eta^{\mu\nu}\X_\nu \otimes \M_{12}}e^{-\M_{12} \otimes \zeta^{(\rho)}_{\mu}\eta^{\mu\nu}\X_\nu} \, ,
\end{equation}
and can be used to exchange two plane wave evaluated in different points
\begin{equation}
\label{eq:rhoexchange}
\begin{aligned}
    \E_b[\eta]\E_a[\p]=&\mu\circ R \triangleright \E_a[\p]\otimes \E_b[\eta] = \mu\circ e^{\zeta_\mu^{(\rho)}\eta^{\mu\nu}\X_\nu \otimes \M_{12}}e^{-\M_{12} \otimes \zeta^{(\rho)}_{\mu}\eta^{\mu\nu}\X_\nu} \triangleright \E_a[\p]\otimes \E_b[\eta]=\\
    &\E_a[\p_0,M(-\tilde{\eta})\cdot\Vec{\p}_\perp,\p_3]\E_b[\eta_0,M(\tilde\xi)\cdot\Vec{\eta}_\perp,\eta_3] \, ,
    \end{aligned}
\end{equation}
where the $\perp$ subscript refers to components $1$ and $2$ of the momenta and
\begin{equation}
M(\tilde\xi )=\left(\begin{array}{cc}
        \cos(\tilde\xi ) & \sin(\tilde\xi ) \\
        -\sin(\tilde\xi ) & \cos(\tilde\xi )
    \end{array}\right) \,.
\end{equation}
Notice that the antipode can be written in compact form as $S_\mu(\xi)=(-\xi_0,-M(\tilde\xi)\cdot\vec\xi_\perp,-\xi_3)$ and that $S(\tilde\xi)=-\tilde\xi$. From \eqref{eq:rhoexchange} we can read off the momentum space representation of the $R$-matrix and by means of \eqref{eq:deformedoscillatoralgebra} we can write the deformed oscillator algebra:
\begin{equation}
    \label{eq:rhodho}
    \begin{aligned}
        &a(\vec\p)\,a(\vec\eta)=a(M(\tilde\xi)\cdot\vec{\eta}_\perp,\eta_3)~a(M(-\tilde\eta)\cdot\vec{\p}_\perp,\p_3) \\
        &a^\dagger(\vec\p)\,a^\dagger(\vec\eta)=a^\dagger(M(-\tilde\xi)\cdot\vec{\eta}_\perp,\eta_3)~a^\dagger(M(\tilde\eta)\cdot\vec{\p}_\perp,\p_3)\\
        &a(\vec\p)\,a^\dagger(\vec\eta)-a^\dagger(M(\tilde\xi)\cdot\vec{\eta},\eta_3) ~a(M(\tilde\eta)\cdot\Vec{\p},\p_3)=2\omega_{\vec\p}\delta(\vec{\p}-\vec{\eta})\delta(\pi_3-\eta_3) \, ,
    \end{aligned}
\end{equation}
which can be used to check that
$[\phi^+_a,\phi^-_b]=W(\x_a-\x_b)$ and that $[\phi_a^{\pm},W(\x_b-\x_c)]$, as assured by the results in \Cref{sec:wicktheorem}, resulting in an undeformed version of the Wick theorem, with undeformed $N$-point functions.

\subsection{Introducing interactions}
\label{sec:interactions}

Up to this point, we proved that free scalar field theories on all unimodular T-Minkowski models are indistinguishable from the commutative case. This applies to models 10--16 and 19--21, which include all three $\zeta$-Minkowski models (among which $\varrho$- and $\lambda$-Minkowski), the famous $\theta$-Minkowski, and four other models numbered 10, 11, 12 and 16 in~\cite{Mercati:2023apu}. The Poincaré-violating IR/UV mixing, which is present even in the free theories (in the 4-point function) when a non-covariant quantization scheme is employed, disappears in our formalism. The price to pay is that also all dependence on the noncommutativity scale vanishes from the expression of the N-point functions, leaving a theory that is functionally identical to QFT on commutative Minkowski spacetime.
In this section we investigate whether this result applies also to interacting theories, at least at the perturbative level. We will propose a tentative generalization of the Gell-Mann--Low formula, which applies to all unimodular T-Minkowski models, but for which we do not have a rigorous derivation. A subset of these models, however, have a crucial property, namely that the difference between time coordinates $x_a^0 - x_b^0$ is a central element of $C_\ell[\mathbbm{R}^{3,1}]^{\otimes_\ell N}$. This property allows to follow all the steps of the derivation of the Gell-Mann--Low formula with the same results as in the commutative theory, and therefore the formula seems to be a rigorous consequence of our covariant quantization scheme, in the presence of interactions, for these models. The models are the three $\zeta$-Minkowski and the $\theta$-Minkowski ones, \textit{i.e.} numbers 13--15 and 19--21 in~\cite{Mercati:2023apu}.

Let us start by recalling that in a commutative QFT of a real scalar field with interaction Hamiltonian $\mathcal{H}_{int}$, all interacting $N$-point functions can be deduced  from the Gell-Mann-Low formula
\begin{equation}
\label{eq:commgellmanlow}
\langle \Omega |T\{\phi(x_1)\cdots \phi(x_n)\}|\Omega\rangle 
= \frac{\langle 0|T\{\phi_I(x_1)\cdots\phi_I(x_n)e^{-i\int d^4y\,\mathcal{H}_{\text{int}}[\phi_I(y)]} \}|0\rangle}{\langle 0|T e^{-i\int d^4z\,\mathcal{H}_{\text{int}}[\phi_I(z)]} |0\rangle}.
\end{equation}
Here, $|\Omega\rangle$ is the interacting vacuum, $|0\rangle$ the free vacuum, and fields $\phi_I$ are interaction-picture fields (\textit{i.e.} solutions of the free equations of motion, written in terms of the free oscillator algebra acting on the free Fock space), and $T$ denotes time ordering, which for a pair of fields is defined as
\begin{equation}
\label{eq:commtimeordering}
    T\{\phi(x_1)\phi(x_2)\}=\phi(x_1)\phi(x_2)\theta(x_1^0-x_2^0)+\phi(x_2)\phi(x_1)\theta(x_2^0-x_1^0) \,,
\end{equation}
($\theta(x)$ is the Heaviside function), and is generalized to $N$-fields in the usual way. For definiteness, we focus on non-derivative interactions, so that the integrated interaction Hamiltonian operator appearing in the exponential is none other than the interaction part of the action, \textit{i.e.}
\begin{equation}
    \mathcal{S}_{int}=\int d^4y \,\,\mathcal{H}_{int}[\phi_I(y)] =-\int d^4y \,\,\mathcal{L}_{int}[\phi_I(y)] \,.
\end{equation}
The computation of $N$-point functions in perturbation series is then realized by expanding the exponential of the interacting part of the action in powers of the coupling constant. Notice that the T-ordering operator acts on the time evolution operator, $\exp ( -i\int d^4y\,\mathcal{H}_{\text{int}}[\phi_I(y)])$, as well.

An extrapolation of \eqref{eq:commgellmanlow} to the noncommutative setting requires us to generalize the time ordering operation in a suitable way. The Heaviside theta function can be easily defined as an element of $\mathcal{M}^{N-1}$ in the standard way:
\begin{equation}
    \theta(\x^0_a - \x^0_b) = \frac 1 {2\pi \, i} \lim_{\epsilon \to 0} \int_{-\infty}^\infty d \tau \, \frac{e^{i \, \tau (\x^0_a - \x^0_b)}}{\tau + i \epsilon} \,,
\end{equation}
because it is a function of coordinate differences, which close the Abelian subalgebra $\mathcal{M}^{N-1}$. However, such a function will not commute, in general, with noncommutative fields, because of the semidirect product structure of $ C_\ell[\mathbbm{R}^{3,1}]^{\otimes_\ell N} =  \mathcal{M}^{cm}_\ell \ltimes \mathcal{M}^{N-1}$. In fact, consider the left multiplication by a plane wave. Eq.~\eqref{eq:th1} implies:
\begin{equation}
    \E_a[p] \, \theta(\x^0_b - \x^0_c)  = \theta \left[ (\chi^{-1})^0{}_\mu(p) (\x^\mu_b - \x^\mu_c) \right]  \, \E_a[p] \,,
\end{equation}
and, as is well-known, since $(\chi^{-1})^0{}_\mu(p)$ is a Lorentz matrix, the expression $(\chi^{-1})^0{}_\mu(p) (\x^\mu_b - \x^\mu_c)$ has the same sign of $(\x^0_b - \x^0_c)$ only if $(\x^\mu_b - \x^\mu_c)$ is timelike. In other words:
\begin{equation}
\x^\mu_b - \x^\mu_c ~\text{ timelike} ~~ \Rightarrow ~~ \left[  \E_a[p] , \theta(\x^0_b - \x^0_c)   \right] = 0  \,,
\end{equation}
but if $(\x^\mu_b - \x^\mu_c)$ is spacelike, $\E_a[p]$ and  $\theta(\x^0_b - \x^0_c) $ might not commute.

As anticipated, we can identify two classes of models in which the T-ordering can be defined unambiguously: according to Zakrzewski's numbering adopted in in~\cite{Mercati:2023apu,Mercati:2024rzg}, they are cases 13 to 15 (the so-called ``$\zeta$-Minkowski'' models, including $\varrho$- and $\lambda$-Minkowski) and cases 19 to 21 (Moyal/$\theta$-Minkowski spacetimes). As shown in Appendix~\ref{app:Rmatrices}, in these models the Lorentz matrix $\chi^\mu{}_\nu(p)$ is such that $(\chi^{-1})^0{}_\mu(p) = 0$ for any value of $p$, and so the time components of the coordinate differences are central elements of $ C_\ell[\mathbbm{R}^{3,1}]^{\otimes_\ell N} $, and commute with everything. This allows to introduce a time ordering operator acting on fields, and to generalize straightforwardly the Gell-Mann--Low formula~\eqref{eq:commgellmanlow}:
\begin{equation}
\label{eq:noncommgellmanlow}
\langle \Omega |T\{\phi(\x_1)\cdots \phi(\x_n)\}|\Omega\rangle 
= \frac{\langle 0| \mathcal{T}\{\phi_I(\x_1)\cdots\phi_I(\x_n)e^{-i\int d^4\y\,\mathcal{H}_{\text{int}}[\phi_I(\y)]} \}|0\rangle}{\langle 0| \mathcal{T} e^{-i\int d^4\hat{z}\,\mathcal{H}_{\text{int}}[\phi_I(\hat{z})]} |0\rangle} \,,
\end{equation}
where
\begin{equation}
 \mathcal{T} \triangleright f(\x_1 , \x_2, \dots , \x_N) = \sum_{\pi\in S_N}
\left[
  \prod_{k=1}^{N-1}
  \theta\!\bigl(\x_{\pi(k)}^{0}-\x_{\pi(k+1)}^{0}\bigr)
\right]
\,
f(\x_{\pi(1)},\x_{\pi(2)}, \dots ,\x_{\pi(N)})\,.
\end{equation}
This implies that, for models 13--15 and 19--21, any QFT, free or interacting (at the perturbative level, at least), is undeformed and will yield identical N-point functions to the commutative case. The only condition is that the Lagrangians are built, using the differential-geometric tools developed in \cite{Mercati:2024rzg}, to be invariant under the coaction of the T-Poincaré quantum group. For example, polynomial vertices should only involve the noncommutative product of $C_\ell[\mathbbm{R}^{3,1}]$, like
\begin{equation}
    \phi^4(\x) = \phi(\x) \cdot \phi(\x)\cdot \phi(\x)\cdot \phi(\x) = \int d^4\xi_1 \dots d^4\xi_4 \, \tilde{\phi}(\xi_1) \dots  \tilde{\phi}(\xi_4) \E[\Delta(\xi_1,\xi_2,\xi_3 ,\xi_4)] \,.
\end{equation}
The readers more accustomed with  the formalism of star products, might say that one needs to multiply the four fields with the star product, and not the commutative product.
The triviality of field theories on Moyal/$\theta$-Minkowski spacetimes  has already been proven, with exactly the same argument used here, in~\cite{Fiore:2007vg}. The works that derive deformed physical predictions from QFTs on Moyal/$\theta$-Minkowski spacetimes, \textit{e.g.}~\cite{Minwalla:1999px,Hayakawa:1999yt,Gomis:2000zz,Matusis:2000jf,Gubser:2000cd,Carroll:2001ws,Chaichian:2004yh,Chaichian:2004za} do so at the cost of using a formalism that breaks the Relativity Principle. Works that adopt a  covariant quantization scheme, as the already-mentioned~\cite{Fiore:2007vg}, but also~\cite{oeckl2000untwisting} and~\cite{Giotopoulos:2021ieg,DimitrijevicCiric:2023hua}, find undeformed results in complete agreement with ours. In particular, it is worth pointing out that the approach of~\cite{Giotopoulos:2021ieg,DimitrijevicCiric:2023hua}, based on BV quantization and $L_\infty$ algebras, present their N-point functions in Fourier transform, which means that the terms corresponding to non-planar diagrams or non-orientable vertices have momentum-dependent phases that resembles the IR/UV mixing terms of the non-covariant approaches. This might lead the readers to conclude that their predictions are Poincaré-violating, but, in actuality, these terms appear as a consequence of the ``braided Wick theorem'', and cancel out once one takes the inverse Fourier transform of the N-point functions with noncommutative fields, yielding undeformed results when expressed in direct space.

The advancement we presented in this Section is that a proof in the same style as~\cite{Fiore:2007vg} can be applied also to the $\zeta$-Minkowski spacetimes (cases 13 to 15  in~\cite{Mercati:2023apu,Mercati:2024rzg}): $\varrho$-Minkowski, $\lambda$-Minkowski and their lightlike counterpart. The result is that any QFT, interacting or not, built on these noncommutative spacetimes, in order to be relativistically covariant, needs to be trivial (in the sense that its N-point functions do not depend on the noncommutativity parameter and are indistinguishable from their commutative counterparts). In this case too, works like~\cite{DimitrijevicCiric:2018,HersentWallet:2023,Hersent:2024,MarisWallet:2024,Wallet:2025} which find deformed physical predictions, must be doing so at the cost of breaking Poincaré covariance (and breaking both ``standard'' and ``T-Poincaré'', quantum group-based covariance).

Finally, the Gell-Mann--Low formula~\eqref{eq:noncommgellmanlow} cannot be proven at this stage for the remaining four unimodular models, 10--12 and 16, however it seems plausible that an analogue formula, with an appropriate generalization of the T-ordered product, will hold, and allow to calculate all perturbative contributions to the interacting N-point functions.

\subsection{The natural path integral and its problems}

The issues with defining the T-ordering in models other than $\zeta$- and $\theta$-Minkowski motivate the consideration of alternatives for defining interacting QFTs. The most obvious one is the path integral formulation. In the commutative case, (T-ordered, interacting) N-point functions are defined as the expectation value of a product of \textit{classical} fields, calculated at different points, with respect to the (normalized) functional measure provided by the exponential of the action. The field configurations one integrates over are classical (\textit{i.e.} they are not operator-valued), and so the fields in the functional integral commute with each other, and no T-ordering is necessary. The T-ordering of the final product is ensured by defining the path integral over an appropriate integration contour in the complexified momentum space (a procedure that is necessary also to define the N-point function as the boundary value of a complex function in its analytic domain, and give meaning to integrals that would be divergent on the real domain).
All the steps just described can be carried out in the noncommutative case, if one defines a (standard) path integral in Fourier transform. This is natural, because the Fourier transform $\tilde{\phi}(p)$ of a noncommutative field $\phi(\x)$ is a commutative function, and the standard path integral over  $\tilde{\phi}(p)$ can be used as a definition for the path integral over noncommutative fields. In this paper, we will call this choice the \textit{natural} path integral, and we will show that it leads to a breakdown of the T-Poincaré covariance of the theory, necessitating a more advanced notion of path integral. In order to show the T-Poincaré breaking, it is sufficient to consider a free scalar field. If the path integral formulation agreed with the  covariant quantization framework discussed in sec.~\ref{sec:covariantquantization}, then all N-point functions of a theory on any unimodular T-Minkowski spacetime would end up undeformed. We can show that this is not the case, once one considers 4-point functions.

The action of a real free scalar field on a unimodular T-Minkowski model is (recall that, since we specialized to unimodular models, we have dropped the ``left'' and ``right'' pedices distinguishing the two Fourier transforms):
\begin{equation}\label{RealAction}
    S   = {\frac 1 2} \int d^4 \xi \,  \tilde{\phi}(\xi)  \tilde{\phi}[S(\xi)] \, \mathcal{C}(\xi)  \,,
\end{equation}
The functional integral measure will be defined on Fourier transformed fields, so that the N-point functions can be written as:
\begin{equation}
    G_N(\x_1,\x_2,\dots,\x_N)  = 
    \mathcal{N} \, \int d^4\xi_1 \dots d^4\xi_N \, \E_1[\xi_1] \dots \E_2[\xi_N]  \int \mathcal{D}[\tilde{\phi}] \,   \tilde{\phi}(\xi_1) \dots \tilde{\phi}(\xi_N) \,  e^{i S[\tilde{\phi}]} \,, 
\end{equation}
where $\mathcal{N}=1/\int \mathcal{D}[\tilde{\phi}] \exp (i S[\tilde{\phi}])$. The expression above holds if $N$ is even, and is zero when $N$ is odd. The momentum-space path integral satisfies a form of Wick's theorem (in the following, we omit all numerical factors and write proportionality symbols instead):
\begin{equation}
    \mathcal{N} \, \int \mathcal{D}[\tilde{\phi}] \,   \tilde{\phi}(\xi_1) \dots \tilde{\phi}(\xi_N) \,  e^{i S[\tilde{\phi}]} \propto  \sum_{P \in \text{pairings}}
\;\prod_{(i,j)\in P}
\;\frac{i}{\mathcal{C}(\xi_i)+i\epsilon}
\,\delta^{(4)}\!\bigl[\xi_j - S(\xi_i)\bigr]
\end{equation}
and it is almost undeformed, the only difference with its commutative counterparts being the antipode appearing in each delta function.

The above expression implies that the two-point function is undeformed. In fact:
\begin{equation}
    \mathcal{N} \, \int \mathcal{D}[\tilde{\phi}] \,   \tilde{\phi}(\xi_1) \, \tilde{\phi}(\xi_2) \,  e^{i S[\tilde{\phi}]} \propto   
 \frac{i}{\mathcal{C}(\xi_1)+i\epsilon}
\,\delta^{(4)}\!\bigl[\xi_2 - S(\xi_1)\bigr] \,,
\end{equation}
which means
\begin{equation}
     G_2(\x_1,\x_2)  \propto 
i \int d^4\xi_1 \, d^4\xi_2 \, \E_1[\xi_1] \,\E_2[\xi_2] \,  \frac{\delta^{(4)}\!\bigl[\xi_2 - S(\xi_1)\bigr]}{\mathcal{C}(\xi_1)+i\epsilon} \,,
\end{equation}
the antipode in the delta function saves the day: integrating with respect to $\xi_2$:
\begin{equation}
     G_2(\x_1,\x_2)  \propto
i \int d^4\xi_1 \frac{\E_1[\xi_1] \,\E_2^*[\xi_1]}{\mathcal{C}(\xi_1)+i\epsilon} \propto
i \, \int d^4\xi \,  \frac{e^{i \, \xi_\mu (\x^\mu_1 - \x^\mu_2)}} {\xi_\mu \xi^\mu - m^2 +i\epsilon} \propto \Delta_\st{F}(\x_1-\x_2) \,,
\end{equation}
which is the undeformed Feynman propagator.
So far, there is agreement with the quantization of sec.~\ref{sec:covariantquantization}. The four-point function, however, works differently. In fact, 
\begin{align}
        &\mathcal{N} \, \int \mathcal{D}[\tilde{\phi}] \,   \tilde{\phi}(\xi_1) \dots \tilde{\phi}(\xi_4) \,  e^{i S[\tilde{\phi}]} \propto 
%     \Big[
%      \frac{\delta^{(4)}[S(\xi_1)-\xi_2]\delta^{(4)}[S(\xi_3)-\xi_4]}{(\mathcal{C}(\xi_1)+i\epsilon)(\mathcal{C}(\xi_3)+i\epsilon)}
% \\
% &+\frac{\delta^{(4)}[S(\xi_1)-\xi_3]\delta^{(4)}[S(\xi_2)-\xi_4]}{(\mathcal{C}(\xi_1)+i\epsilon)(\mathcal{C}(\xi_2)+i\epsilon)}
% \\
% &+\frac{\delta^{(4)}[S(\xi_1)-\xi_4]\delta^{(4)}[S(\xi_2)-\xi_3]}{(\mathcal{C}(\xi_1)+i\epsilon)(\mathcal{C}(\xi_2)+i\epsilon)}
% \Big] \,,
\\
&\frac{\delta^{(4)}[S(\xi_1)-\xi_2]\delta^{(4)}[S(\xi_3)-\xi_4]}{(\mathcal{C}(\xi_1)+i\epsilon)(\mathcal{C}(\xi_3)+i\epsilon)}
+\frac{\delta^{(4)}[S(\xi_1)-\xi_3]\delta^{(4)}[S(\xi_2)-\xi_4]}{(\mathcal{C}(\xi_1)+i\epsilon)(\mathcal{C}(\xi_2)+i\epsilon)}
+\frac{\delta^{(4)}[S(\xi_1)-\xi_4]\delta^{(4)}[S(\xi_2)-\xi_3]}{(\mathcal{C}(\xi_1)+i\epsilon)(\mathcal{C}(\xi_2)+i\epsilon)} \,,
\nonumber
\end{align}
then,
\begin{equation}
\begin{aligned}
&G_4(\x_1,\x_2,\x_3,\x_4) \propto 
\int \frac{d^4\xi_1 d^4 \xi_2}{[\mathcal{C}(\xi_1)+i\epsilon][\mathcal{C}(\xi_2)+i\epsilon]}  \times
\\
&\times \Big(
\E_1[\xi_1] \, \E_2^*[\xi_1]\, \E_3[\xi_2]\, \E_4^*[\xi_2]  + \E_1[\xi_1] \, \E_2[\xi_2]\, \E_3^*[\xi_2]\, \E_4^*[\xi_1]
+\E_1[\xi_1] \, \E_2[\xi_2]\, \E_3^*[\xi_1]\, \E_4^*[\xi_2] 
\Big)\,,
\end{aligned}
\end{equation}
the first and second terms are undeformed:
\begin{equation}
    \begin{aligned}
        &\E_1[\xi_1] \, \E_2^*[\xi_1]\, \E_3[\xi_2]\, \E_4^*[\xi_2] = e^{i \, \xi_1 \cdot (x_1-x_2)}\, e^{i \, \xi_2 \cdot (x_3-x_4)} \,, \\
        &\E_1[\xi_1] \, \E_2[\xi_2]\, \E_3^*[\xi_2]\, \E_4^*[\xi_1] = e^{i \, \xi_1 \cdot (x_1-x_4)}\, e^{i \, \xi_2 \cdot (x_2-x_3)} \,,
    \end{aligned}
\end{equation}
while the third is deformed, and not translation-invariant:
\begin{equation}\label{eq:ThirdTerm}
    \E_1[\xi_1] \, \E_2[\xi_2]\, \E_3^*[\xi_1]\, \E_4^*[\xi_2]
    = \E_1[\xi_1]\, \E_3^*[R^{(1)}(\xi_2,\xi_1)] \, \E_2[R^{(2)}(\xi_2,\xi_1)]\, \E_4^*[\xi_2] \,.
\end{equation}
This proves that the ``natural'' path integral is not equivalent to the quantization of sec.~\ref{sec:covariantquantization}, and indeed it breaks T-Poincaré invariance. The same results would be obtained by calculating the free N-point functions for noncommutative quantum fields whose oscillator algebra is undeformed, in place of~\eqref{eq:deformedoscillatoralgebra}. 
This matches with what found in~\cite{oeckl2000untwisting} for $\theta$-Minkowski: one can deform the product between functions (\textit{i.e.} make it noncommutative) and the oscillator algebra. Doing only one of the two things leads to Poincaré-violating results, which are also affected by IR/UV mixing. Deforming both spacetime and oscillator algebras leads to Poincaré-invariant results, which however are identical to the ones of the undeformed/commutative theory, in which both spacetime and the oscillator algebra are undeformed.

The covariant quantization scheme presented in sec.~\ref{sec:covariantquantization} corresponds, at least at the level of physical predictions, to the \textit{braided QFT} introduced by Oeckl~\cite{Oeckl:1999zu}.\footnote{Which is also the quantization scheme behind the approach of~\cite{Giotopoulos:2021ieg,DimitrijevicCiric:2023hua}.} This approach allows to derive the N-point functions from a path integral, but ensures that the correct momentum-dependent terms are pulled down when a sequence of functional derivatives are tangled in a nontrivial way [as in the third term above, Eq.~\eqref{eq:ThirdTerm}], producing a deformed Wick theorem which makes the N-point functions Poincaré invariant when expressed in direct space.

\section{Outlook and conclusions}

Building upon the mathematical formalism of T-Minkowski noncommutative spacetimes developed in \cite{Mercati:2023apu,Mercati:2024rzg}, we were able to construct the Quantum Field Theory of a real free scalar field, covariant under T-Poincaré transformations, for any of the 10 unimodular models (10--16 and 19--21 in~\cite{Mercati:2023apu}). The  result was extended rigorously to interacting scalar fields (at the perturbative level) only in the case of the six ``$\zeta$'' and ``$\theta$'' models (13--15 and 19--21 in~\cite{Mercati:2023apu}), even though it plausibly extends to the remaining four unimodular models.   The key ingredient of the construction is an appropriate non-commutative generalization of the traditional covariant quantization scheme of communtative QFT (as described, for example, in \cite{schweber2013introduction}). In the commutative case, the prescription requires that the scalar field satisfies the Klein--Gordon equation and that the commutator between fields at two spacetime points is a Poincaré invariant function. It can be shown that these requirements constrain the commutator to be proportional to the Pauli-Jordan function. In our non-commutative setting, we simply require that the scalar field satisfies the non-commutative Klein--Gordon equation and that the commutator between fields living in different subalgebras of the braided tensor product algebra is a non-commutative function $f\in C_\ell[\mathbbm{R}^{3,1}]^{\otimes_\ell N}$ invariant under T-Poincaré transformations. We obtain that $f$ depends solely on coordinate differences, which are commutative, and actually coincides with the commutative Pauli-Jordan function. By restricting to unimodular cases, so that we can expand the field into non-commutative generalizations of positive and negative frequency modes, we are able to prove a Wick theorem for the $N$-point Wightmann functions of the free theory. The result is structurally equivalent to the one found in the commutative case, and all $N$-point Wightmann functions coincide with their commutative counterpart.
The mechanism behind the cancellation of non-commutative effects is analogous to the one identified in Ref. \cite{Fiore:2007vg} for the case of $\theta$ non-commutativity. As a consequence of requiring covariance of the QFT under T-Poincaré, the deformations arising from the non-commutativity of plane waves living in different subalgebras of $C_\ell[\mathbbm{R}^{3,1}]$ are reabsorbed by the deformations in the oscillator algebra, as observed in~\cite{oeckl2000untwisting} for the case of $\theta$-Minkowski.   In cases 13 to 15 (the so called ``$\zeta$-Minkowski'') and cases 19 to 21 ($\theta$-Minkowski), time coordinate differences are central elements of $C_\ell[\mathbbm{R}^{3,1}]^{\otimes_\ell N}$, which allows us to propose a generalization of the Gell--Mann--Low formula and conclude that the $N$-point functions for interacting theories all coincide with their commutative counterparts. Most notably, this implies that these models are unaffected by IR/UV mixing, contrary to the findings of different approaches~\cite{DimitrijevicCiric:2018,HersentWallet:2023,Hersent:2024,MarisWallet:2024,Wallet:2025} which forfeit relativistic covariance.

In closing, we observe that the most urgent future task would be to introduce interactions for the remaining unimodular models. We conjecture that this could be achieved by implementing interactions in a covariant way, avoiding the standard Hamiltonian treatment relying on the commutativity of time coordinate differences. The most promising path towards this goal seems to be the prescriptions of Oeckl's braided path integral~\cite{Oeckl:1999zu}, designed to match the effects of the deformed harmonic oscillator algebra. In turn, this could aid the computation of $N$-point functions for non-unimodular models as well. So far, the only example of a covariant noncommutative QFT for these models is the one developed on lightlike $\kappa$-Minkowski \cite{Lizzi:2021rlb,DiLuca:2022idu,Fabiano:2023xke}, where only the 2-point function is computed for the free theory.
Braided QFT relies on the existence (and explicit calculability) of infinite-dimensional representations of the R-matrix, and all T-Minkowki models admit a universal R-matrix that fits the description (we calculate them explicitly in~\cref{app:Rmatrices}). However, non-unimodular models all have a \textit{singular} R-matrix, with terms that diverge at certain values of the momenta. This can be solved by using Lorentz-covariant momentum coordinates, but the map from the original momentum space to these coordinates is not one-to-one. This signals the necessity to introduce a second region of momentum space which is not connected under the momentum composition law to the first, while Lorentz transformations do connect the two regions. This was done in the lightlike $\kappa$-Minkowski model in~\cite{DiLuca:2022idu,Fabiano:2023xke}, and can be expected to be feasible in a similar way for all non-unimodular T-Minkowski models.

% It may also be interesting to extend our framework to non‑Abelian gauge theories on T‑Minkowski, as the noncommutativity of the gauge group could have some interplay with that of the coordinates, as observed in \cite{}. \textcolor{red}{sbaglio o c'erano dei lavori di Patrizia che parlavano di questo? Ce ne è uno mio e di Angel! Patrizia non so. Cmq non lo menzioniamo}

\section*{Acknowledgements}
This work contributes to the European Union COST Action CA23130 \emph{Bridging high and low energies in search of quantum gravity}.

F.M. acknowledges support by the Agencia Estatal de Investigación (Spain)
under grants CNS2023-143760 and PID2023-148373NB-I00 funded by MCIN/AEI/10.13039/501100011033/FEDER – UE,
and by the Q-CAYLE Project funded by the Regional Government of Castilla y León (Junta de Castilla y León) and by the Ministry of Science and Innovation MICIN through NextGenerationEU (PRTR C17.I1). G.F. thanks the Universidad de Burgos for their hospitality and acknowledges financial support from ``Fondazione Angelo Della Riccia'' and from the ``Foundation Blanceflor''.

\newpage 

\appendix

\section{Unimodular T-Minkowski models and their properties}
\label{app:Rmatrices}
In this appendix we report useful properties of all the unimodular T-Minkowski models, where we set to 0 the Moyal central extension for the ones with non-zero Lie algebriac structure. We write  the explicit expressions for $f^{\mu\nu}{}{}_\rho$ and the ensuing coordinate commutation relations. For each model, we choose a convenient prescription for the ordering of plane waves and derive the expressions for the linear momentum coordinates $\xi_\mu$ and the relevant $SO(3,1)$ matrix $\chi^\mu{}_\nu$, by means of \eqref{Eq:ChiXiRepresentation}.
From the explicit expressions of $\xi_\mu(p)$ one is able to show that for every model, $\xi_\mu(p)$ is a bijection of $\mathbbm{R}^4$. We also derive the antipode in linear momentum coordinates using  $S(\xi)_\mu=-\xi_\nu \,\chi^\nu{}_\mu(\xi)$ (which is consequence of \eqref{ExteriorDifferentialExplicitForm} and the fact that $\chi^\mu{}_\nu(p)$ is a $SO(3,1)$ matrix) and show that it satisfies the properties outlined in the beginning \cref{sec:wicktheorem}, required to conveniently expand the scalar field in terms of non-commutative generalizations of positive and negative frequency plane waves.
Finally, we report 
the expression for the $r$-matrix which, following the prescription in \cite{tolstoy2007twistedquantumdeformationslorentz}, allows us to compute the twist element and, consequently the $R$-matrix, through the relation
\begin{equation}
   \mathcal{R} = \mathcal{F}_{21}^{-1} \, \mathcal{F}_{12} \,,
\end{equation}
as shown in \cref{sec:rmat}.
This enables us to derive the momentum space representation of the $R$-matrix, needed for the proof in \cref{app:oscillatoralgebra}.

The conventions for the R-matrices are chosen such that 
\begin{equation}
    \x^\mu\x^\nu=\mu\circ R\triangleright \x^\nu\otimes \x^\mu
\end{equation}
The action of $\X_\mu$ and $\M_{\rho\sigma}$ on $\x^\mu$ can be deduced using equations \eqref{ExteriorDifferentialExplicitForm} and \eqref{eq:linPoinc}, respectively,yielding
\begin{equation}
    \X_\mu\triangleright \x^\nu=-i\delta^\mu_\mu \, , \qquad \M_{\rho\sigma}\triangleright \x^\mu=(\eta_{\rho\nu}\delta^\mu_\sigma -\eta_{\sigma\nu}\delta^\mu_\rho)\x^\nu \, .
\end{equation}

\begin{itemize}

\item \textbf{Case 10}

\begin{equation}
    \begin{aligned}
    r_{10} =~&  \mP_1 \wedge (\mM_{23}-\mM_{02})+(\mP_0 + \mP_3) \wedge (\mM_{01}-\mM_{13}) \,,
\\
f^{\mu \nu }{}_{\rho } =&- \left(\delta ^{\nu }{}_0+\delta ^{\nu }{}_3\right) \left(\eta _{\rho 1} \delta ^{\mu }{}_0-\eta _{\rho 0} \delta ^{\mu }{}_1-\eta _{\rho 3} \delta ^{\mu }{}_1+\eta _{\rho 1} \delta ^{\mu }{}_3\right)
\\
&-\delta ^{\nu }{}_1 \left(-\eta _{\rho 2} \delta ^{\mu }{}_0+\eta _{\rho 0} \delta ^{\mu }{}_2+\eta _{\rho 3} \delta ^{\mu }{}_2-\eta _{\rho 2} \delta ^{\mu }{}_3\right)\,,
\end{aligned}
\end{equation}
\begin{equation}
   \begin{aligned}
& [\x^0,\x^1]=i \left(\x^3-\x^0-\x^2\right)\,,& &[\x^0,\x^2]=0\,,
 \\ 
 & [\x^1,\x^2]=i \left(\x^0-\x^3\right)\,,& 
 & [\x^0,\x^3]=0 \,,
 \\ 
 & [\x^1,\x^3]=i \left(\x^0+\x^2-\x^3\right)& &[\x^2,\x^3]=0\,.
   \end{aligned}
\end{equation}
To simplify expressions in this case, it will be convenient to adopt light-cone coordinates defined by $p_+=\frac{p_0+p_3}{2}$, $p_-=\frac{p_0-p_3}{2}$, $\x_+=\x_0+\x_3$, $\x_-=\x_0-\x_3$, $\xi_+=\frac{\xi_0+\xi_3}{2},\xi_-=\frac{\xi_0-\xi_3}{2}$ and to choose the ordering prescription
\begin{equation}
    \E[p]=e^{ip_2\x^2}e^{ip_-\x^-}e^{ip_+\x^+}e^{ip_1\x^1}
\end{equation}

\begin{equation}
    \chi^\mu{}_\nu(p)=\left(
\begin{array}{ccccc}
 1+\frac{1}{2}p_1^2+2p_+^2 & -2p_+ & p_1 & -\frac{p_1^2}{2}-2p_+^2 \\
 -2p_+ & 1 & 0 & 2p_+ \\
 p_1 & 0 & 1 & -p_1 \\
 \frac{p_1^2}{2} +2p_+^2 & -2p_+ & p_1 & 1-\frac{p_1^2}{2} -2p_+^2   \\
\end{array}
\right)
\end{equation}
\begin{equation}
\begin{aligned}
    &\xi_+=p_+ \qquad \xi_-=p_-+2p_1p_++\frac{4p_+^3}{3} \qquad \xi_1=p_1+2p_+^2 \qquad \xi_2=p_2\\
    &S(\xi)_+=-\xi_+ \qquad S(\xi)_-=-\xi_--(\xi_1-2\xi_+^2)(\xi_2+\xi_+(\xi_1-2-2\xi_+^2))\\
    &S(\xi)_1=-\xi_1+4\xi_+^2 \qquad S(\xi)_2=-\xi_2-2\xi_1\xi_++4\xi_+^3
\end{aligned}
\end{equation}
\begin{equation}
\begin{aligned}
    \mathcal{R}=&e^{-(\M_{01}-\M_{13})\otimes \X_+}e^{\X_+\otimes(\M_{01}-\M_{13})}e^{-(\M_{23}-\M_{02})\otimes (\X_1-\X_+^2/2)}e^{(\X_1-\X_+^2/2)\otimes(\M_{23}-\M_{02})}\times\\
    \times& e^{-(\M_{01}-\M_{13})\otimes \X_+}e^{\X_+\otimes (\M_{01}-\M_{13})}
    \end{aligned}
\end{equation}

\item \textbf{Case 11}
\begin{equation}
    \begin{aligned}
    r_{11} =& \mP_2 \wedge ( \mM_{01} - \mM_{13} )   \,,
\\
f^{\mu \nu }{}_{\rho } =&-\delta ^{\nu }{}_2 \left(\eta _{\rho 1} \delta ^{\mu }{}_0-\eta _{\rho 0} \delta ^{\mu }{}_1-\eta _{\rho 3} \delta ^{\mu }{}_1+\eta _{\rho 1} \delta ^{\mu }{}_3\right)\,,
\end{aligned}
\end{equation}
\begin{equation}
   \begin{aligned}
& [\x^0,\x^1]=0\,,& &[\x^0,\x^2]=i \x^1\,,& &[\x^1,\x^2]=i \left(\x^0-\x^3\right)\,,
 \\ 
 & [\x^0,\x^3]=0& &[\x^1,\x^3]=0& &[\x^2,\x^3]=-i \x^1\,.
   \end{aligned}
\end{equation}
We choose the $\x^2$ to the right ordering
\begin{equation}
    \E[p]=e^{i(p_0\x^0+p_1\x^1+p_3\x^3)}e^{ip_2\x^2}
\end{equation}
\begin{equation}
\chi^\mu{}_\nu(p)=    \left(
\begin{array}{ccccc}
 \frac{1}{2} \left(p_2^2+2\right) & -p_2 & 0 & -\frac{p_2^2}{2} \\
 -p_2 & 1 & 0 & p_2 \\
 0 & 0 & 1 & 0 \\
 \frac{p_2^2}{2} & -p_2 & 0 & 1-\frac{p_2^2}{2} \\
\end{array}
\right)
\end{equation}
\begin{equation}
    \begin{aligned}
        &\xi_\mu=p_\mu\\
        &S(\xi)_\mu=(-\p_0-\frac{\p_2^2}{2}(\p_0+\p_3)+\p_1\p_2,-\p_1+\p_2(\p_0+\p_3),-\p_2,-\p_3+\frac{\p_2^2}{2}(\p_0+\p_3)-\p_1\p_2)
    \end{aligned}
\end{equation}
\begin{equation}
    \mathcal{R}=e^{\X_2\otimes (\M_{01}-\M_{13})}e^{-(\M_{01}-\M_{13})\otimes \X_2}
\end{equation}

\item \textbf{Case 12}\footnote{There is another  mistake in~\cite{Zakrzewski_1997}. Case 12 in Table 1 has an additional term, of the form $(\mP_0-\mP_3) \wedge \mP_2$, which does not satisfy the CYBE. This term has been removed here. This mistake was pointed out in~\cite{tolstoy2007twisted} (pag.~15) first, and more recently in~\cite{Meier:2023kzt,Meier:2023lku}.}
\begin{equation}
\begin{aligned}
    r_{12} =&  (\mP_0+\mP_3) \wedge  \left( \mM_{01} - \mM_{13}   \right) \,,
\\
f^{\mu \nu }{}_{\rho } =&-\left( \delta ^{\nu }{}_0+\delta ^{\nu }{}_3\right) \left(\eta _{\rho 1} \delta ^{\mu }{}_0-\eta _{\rho 0} \delta ^{\mu }{}_1-\eta _{\rho 3} \delta ^{\mu }{}_1+\eta _{\rho 1} \delta ^{\mu }{}_3\right)\,,
\end{aligned}
\end{equation}

\begin{equation}
   \begin{aligned}
        & [\x^0,\x^1]=-i \left(\x^0-\x^3\right)\,,& &[\x^0,\x^2]=0\,,& &[\x^1,\x^2]=0\,,
 \\ 
 & [\x^0,\x^3]=0& &[\x^1,\x^3]=i \left(\x^0-\x^3\right)& &[\x^2,\x^3]=0\,.
   \end{aligned}
\end{equation}
Here it will also be convenient to adopt light-cone coordinates and the $\x^+$ to the right ordering
\begin{equation}
    \E[p]=e^{ip_-\x^-}e^{ip_1\x^1}e^{ip_2\x^2}e^{ip_+\x^+}
\end{equation}
\begin{equation}
    \chi^\mu{}_\nu(p)=\left(
\begin{array}{ccccc}
  1+2p_+^2 & -2p_+ & 0 & -2p_+^2  \\
 -2p_+ & 1 & 0 & 2p_+ \\
 0 & 0 & 1 & 0 \\
 2p_+^2 & -2p_+ & 0 & 1-2p_+^2  \\
\end{array}
\right)
\end{equation}
\begin{equation}
\begin{aligned}
    &\xi_+=p_+ \qquad \xi_-=p_-+\frac{4p_+^3}{3} \qquad \xi_1=p_1+2p_+^2 \qquad \xi_2=p_2\\
    &S(\xi)_+=-\xi_+ \qquad S(\xi)_-=-\xi_-+2\xi_1\xi_+-4\xi_+^3\\
    &S(\xi)_1=-\xi_1+4\xi_+^2 \qquad S(\xi)_2=-\xi_2
\end{aligned}
\end{equation}

\begin{equation}
    \mathcal{R}=e^{2\X_+\otimes (\M_{01}-\M_{13})}e^{-2(\M_{01}-\M_{13})\otimes \X_+}
\end{equation}

\item \textbf{Case 13, 14 \& 15}

Assume $\zeta^{(\varrho)}_1 = \zeta^{(\varrho)}_2 = 0$. When $\zeta^{(\varrho)}_\mu = \delta^0_\mu$, $\zeta^{(\varrho)}_\mu = \delta^0_\mu + \delta^3{}_\mu$ and  $\zeta^{(\varrho)}_\mu = \delta^3{}_\mu$ respectively, these are the timelike, lightlike and spacelike version of $\varrho$-Poincar\'e~\cite{Lizzi:2022hcq,Fabiano:2023uhg}. This class of models was introduced in~\cite{Lukierski:2005fc}.
\begin{equation}
\begin{aligned}
    r_{\varrho} =&   \zeta^{(\varrho)}_\mu \, \eta^{\mu\nu} \, \mP_\nu \wedge  \mM_{12}    \,,
\\
f^{\mu \nu }{}_{\rho } =&\left(\eta _{\rho 1} \delta ^{\mu }{}_2-\eta _{\rho 2} \delta ^{\mu }{}_1\right) \left(-\delta^\nu_0 \zeta _0{}^{(\varrho )}+\delta^\nu_3 \zeta _3{}^{(\varrho )}\right)\,,
\end{aligned}
\end{equation}
\begin{equation}
   \begin{aligned}
& [\x^0,\x^1]=i \x^2 \zeta _0{}^{(\varrho )}\,,& &[\x^0,\x^2]=-i \x^1 \zeta _0{}^{(\varrho )}\,,& &[\x^1,\x^2]=0\,,
 \\ 
 & [\x^0,\x^3]=0\,,& &[\x^1,\x^3]=i \x^2 \zeta _3{}^{(\varrho )}\,,& &[\x^2,\x^3]=-i \x^1 \zeta _3{}^{(\varrho )}\,.
   \end{aligned}
\end{equation}
For all 3 models, we choose the $\x^0$ to the right ordering
\begin{equation}
    \E[p]=e^{ip_i\x^i}e^{ip_0\x^0} \, .
\end{equation}
For timelike $\varrho$-Minkowski, we have
\begin{equation}
    \chi^\mu{}_\nu(p)=\left(
\begin{array}{ccccc}
 1 & 0 & 0 & 0  \\
 0 & \cos (p_0) & \sin (p_0) & 0  \\
 0 & -\sin (p_0) & \cos (p_0) & 0  \\
 0 & 0 & 0 & 1  \\
\end{array}
\right)
\end{equation}

\begin{equation}
\begin{aligned}
    &\xi_\mu=p_\mu \\
    &S(\xi)_\mu=(-\xi_0,-\xi_1\cos(\xi_0)+\xi_2\sin(\xi_0),-\xi_2\cos(\xi_0)-\xi_1\sin(\xi_0),-\xi_3)
\end{aligned}
\end{equation}
\begin{equation}
    \mathcal{R}=e^{-\X_0\otimes \M_{12}}e^{\M_{12}\otimes \X_0}
\end{equation}
For lightlike $\varrho$-Minkowski, we have
\begin{equation}
   \chi^\mu{}_\nu(p)= \left(
\begin{array}{ccccc}
 1 & 0 & 0 & 0 \\
 0 & \cos (p_0-p_3) & \sin (p_0-p_3) & 0  \\
 0 & -\sin (p_0-p_3) & \cos (p_0-p_3) & 0  \\
 0 & 0 & 0 & 1  \\
\end{array}
\right)
\end{equation}
\begin{equation}
\begin{aligned}
    &\xi_\mu=p_\mu \\
    &S(\xi)_\mu=(-\xi_0,-\xi_1\cos(\xi_0-\xi_3)+\xi_2\sin(\xi_0-\xi_3),-\xi_2\cos(\xi_0-\xi_3)-\xi_1\sin(\xi_0-\xi_3),-\xi_3)
\end{aligned}
\end{equation}
\begin{equation}
    \mathcal{R}=e^{-(\X_0-\X_3)\otimes \M_{12}}e^{ \M_{12}\otimes (\X_0-\X_3)}
\end{equation}
For spacelike $\varrho$-Minkowski, we have
\begin{equation}
    \chi^\mu{}_\nu(p)=\left(
\begin{array}{ccccc}
 1 & 0 & 0 & 0  \\
 0 & \cos (p_3) & -\sin (p_3) & 0 \\
 0 & \sin (p_3) & \cos (p_3) & 0  \\
 0 & 0 & 0 & 1  \\
\end{array}
\right)
\end{equation}
\begin{equation}
\begin{aligned}
    &\xi_\mu=p_\mu \\
    &S(\xi)_\mu=(-\xi_0,-\xi_1\cos(\xi_3)-\xi_2\sin(\xi_3),-\xi_2\cos(\xi_3)+\xi_1\sin(\xi_3),-\xi_3)
\end{aligned}
\end{equation}
\begin{equation}
    \mathcal{R}=e^{\X_3\otimes \M_{12}}e^{- \M_{12}\otimes \X_3}
\end{equation}

\item \textbf{Case 16}
\begin{equation}
\begin{aligned}
    r_{16} =&\mP_1 \wedge  \mM_{03}   \,,
\\
f^{\mu \nu }{}_{\rho } =-&\delta^{\nu}_1 \left(\eta _{\rho 0} \delta ^{\mu }{}_3-\eta _{\rho 3} \delta ^{\mu }{}_0\right)\,,
\end{aligned}
\end{equation}

\begin{equation}
   \begin{aligned}
& [\x^0,\x^1]=i \x^3\,,& &[\x^0,\x^2]=0\,,& &[\x^1,\x^2]=0\,,
 \\ 
 & [\x^0,\x^3]=0\,,& &[\x^1,\x^3]=-i \x^0\,,& &[\x^2,\x^3]=0\,.
   \end{aligned}
\end{equation}
We choose the $\x^1$ to the right ordering 
\begin{equation}
    \E[p]=e^{ip_0\x^0}e^{ip_2\x^2}e^{ip_3\x^3}e^{ip_1\x^1}
\end{equation}
\begin{equation}
    \chi^\mu{}_\nu(p)=\left(
\begin{array}{ccccc}
 \cosh (p_1) & 0 & 0 & \sinh (p_1) \\
 0 & 1 & 0 & 0 \\
 0 & 0 & 1 & 0  \\
 \sinh (p_1) & 0 & 0 & \cosh (p_1)  \\
\end{array}
\right)
\end{equation}
\begin{equation}
\begin{aligned}
    &\xi_\mu=p_\mu\\
    &S(\xi)_\mu=(-p_0\cosh(p_1)-p_3\sinh(p_1),-p_1,-p_2,-p_3\cosh(p_1)-p_0\sinh(p_1))
\end{aligned}
\end{equation}

\begin{equation}
    \mathcal{R}=e^{\X_1\otimes \M_{03}}e^{-\M_{03}\otimes \X_1}
\end{equation}

\item \textbf{Cases 19, 20 \& 21}

These are the three classes of the $\theta$-Poincar\'e group, corresponding to choices of the $\theta^{\mu\nu}$ matrix that are inequivalent under automorphisms of the Poincar\'e group.
\begin{equation}
    r_{\theta} =  - {\sfrac 1 2} \, \theta^{\mu\nu} \,  \mP_\mu \wedge \mP_\nu      \,,
\qquad
 \theta^{\mu\nu} =   \text{\textit{arbitrary}} \,,
\qquad
    \omega^{\mu\nu\rho} = f^{\mu\nu}{}_\rho  = 0 \,,
\end{equation}
\begin{equation}
[\x^\mu ,\x^\nu ]  = i \, \theta^{\mu\nu} \,.
\end{equation}
The differential calculus structures are all undeformed and the momentum space coordinates defined through Weyl ordering 
\begin{equation}
    \E_a[\xi]=e^{i\xi_\mu\x^\mu}
\end{equation}
correspond to linear momentum coordinates. The R-matrix is given by
\begin{equation}
    \mathcal{R}=e^{i\theta^{\mu\nu}\X_\mu\otimes\X_\nu} \, .
\end{equation}
The development of QFT of a real scalar field on this noncommutative spacetime was extensively discussed in \cite{Fiore:2007vg} and reviewed in \cref{sec:theta}.

\end{itemize}

\section{Computation of the deformed oscillator algebra}
\label{app:oscillatoralgebra}
We present the details of the computation leading to the deformed harmonic oscillator algebra in \eqref{eq:deformedoscillatoralgebra}. The idea is to expand the equality $[\phi(\x_a),\phi(\x_b)]=\Delta_{PJ}(\x_a-\x_b)$ and compare the coefficients of the same products of plane waves on both sides of the equation. We have 
\begin{equation}
\label{eq:explicitexpansion}
    \begin{aligned}
        &[\phi(\x_a),\phi(\x_b)]=\\
        =&\int \frac{d^3\xi \, d^3\eta}{4\omega_{\vec \xi}\omega_{\vec \eta}}\left(a(\vec \xi)a(\vec \eta)\e_a^*[\vec \xi]\e_b^*[\vec \eta]+a(\vec \xi)a^\dagger (\vec \eta)\e_a^*[\vec \xi]\e_b[\vec \eta]+a^\dagger(\vec \xi)a(\vec \eta)\e_a[\vec \xi]\e_b^*[\vec \eta]+a^\dagger(\vec \xi)a^\dagger(\vec \eta)\e_a[\vec \xi]\e_b[\vec \eta] \right)+\\
        -&\int \frac{d^3\xi \, d^3\eta}{4\omega_{\vec \xi}\omega_{\vec \eta}}\left(a(\vec \eta)a(\vec \xi)\e_b^*[\vec \eta]\e_a^*[\vec \xi]+a(\vec \eta)a^\dagger (\vec \xi)\e_b^*[\vec \eta]\e_a[\vec \xi]+a^\dagger(\vec \eta)a(\vec \xi)\e_b[\vec \eta]\e_a^*[\vec \xi]+a^\dagger(\vec \eta)a^\dagger(\vec \xi)\e_b[\vec \eta]\e_a[\vec \xi] \right)=\\
        =&\Delta_{PJ}(\x_a-\x_b)=\int \frac{d^3\xi}{2\omega_{\vec \xi}}\left(\e_b[\vec \xi]\e_a^*[\vec \xi]-\e_a[\vec \xi]\e_b^*[\vec \xi]\right) =\int \frac{d^3\xi}{2\omega_{\vec \xi}}\left(\e_a^*[\vec \xi]\e_b[\vec \xi]-\e_b^*[\vec \xi]\e_a[\vec \xi]\right)\,.
    \end{aligned}
\end{equation}
The last expression for the Pauli-Jordan function can be obtained by performing manipulations similar to those of Eq. \eqref{eq:wightmann2}.
Let us begin by inspecting the terms containing two $a^\dagger$ operators, conveniently rewriting them in terms of off-shell plane waves as
\begin{equation}
\label{eq:expressionadad}
    \int_0^\infty d\xi_0 \int d^3\xi \int_0^\infty d\eta_0 \int d^3\eta\, \delta(\xi^2-m^2)\delta(\eta^2-m^2)\left[a^\dagger(\vec \xi)a^\dagger(\vec \eta)\E_a[\xi]\E_b[\eta]-a^\dagger(\vec \eta)a^\dagger(\vec \xi)\E_b[\eta]\E_a[\xi]\right]
\end{equation}
We can reorder the product $\E_b[\eta]\E_a[\xi]$ by means of \eqref{eq:pwexfinal} and write the second term of this integral as
\begin{equation}
\label{eq:secondterm}
    \int_0^\infty d\xi_0 \int d^3\xi \int_0^\infty d\eta_0 \int d^3\eta\, \delta(\xi^2-m^2)\delta(\eta^2-m^2) a^\dagger(\vec \eta)a^\dagger(\vec \xi) \E_a[R_{(1)}(\xi,\eta)]\E_b[R_{(2)}(\xi,\eta)]
\end{equation}
In this last integral, we perform the change of variables 
\begin{equation}
    \label{eq:cov}
    \begin{cases}
        \xi'=R_{(1)}(\xi,\eta)\\
        \eta'=R_{(2)}(\xi,\eta) 
    \end{cases}\rightarrow \quad  
    \begin{cases}
        \xi=R^{-1}_{(1)}(\xi',\eta')\\
        \eta=R^{-1}_{(2)}(\xi',\eta') 
    \end{cases}
\end{equation}
Using the expressions for the $R$-matrices in \cref{app:Rmatrices}, it is possible to obtain their momentum space representation $(R_{(1)}(\xi,\eta),R_{(2)}(\xi,\eta))$ and to show that the change of variables defined in \eqref{eq:cov} is such that the integration region covered by $\xi,\eta$ is mapped into the same region when changing variables to $\xi',\eta'$. Moreover, the on-shell condition remains invariant, since the momentum space maps $R$ are nonother than Lorentz transformations with momentum dependent rapidity/rotation parameters. Alas, it is also possible to show that for all the unimodular models, the determinant of the Jacobian of the transformation from $\xi,\eta$ to $\xi',\eta'$ is 1.
Therefore, we can write \eqref{eq:expressionadad} as
\begin{equation}
\label{eq:fistruleexpr}
    \int \frac{d^3\xi}{2\omega_\xi}\frac{d^3\eta}{2\omega_\eta}\left(a^\dagger(\vec \xi)a^\dagger(\vec\eta)-a^\dagger(R_{(2)}^{-1}(\xi,\eta))a^\dagger(R_{(1)}^{-1}(\xi,\eta))\right)\e_a[\vec\xi]\e_b[\vec\eta] \, .
\end{equation}
Since in \eqref{eq:explicitexpansion} there are no other terms proportional to $\e_a[\vec\xi]\e_b[\vec\eta]$, \eqref{eq:fistruleexpr} must be 0, which yields the deformed commutation relation for two creation operators: 
\begin{equation}
\label{eq:adad}
a^\dagger(\vec \xi)a^\dagger(\vec \eta)=a^\dagger(R_{(2)}^{-1}(\xi,\eta))a^\dagger(R_{(1)}^{-1}(\xi,\eta)) \, .
\end{equation}
The commutation relation for two annihilation operators can be obtained by taking the hermitian conjugate of \eqref{eq:adad}, yielding
\begin{equation}
    \label{eq:aa}a(\vec \xi)a(\vec\eta)=a(R_{(1)}^{-1}(\eta,\xi))a(R_{(2)}^{-1}(\eta,\xi))
\end{equation}
Using the momentum space representations of the R-matrices reported in \cref{app:Rmatrices}, it is possible to show that \eqref{eq:aa} is consistent with the commutation relation obtained by comparing the terms in \eqref{eq:explicitexpansion} proportional to the products $\e_a^*[\vec \xi]\e_b^*[\vec \eta]$ and $\e_b^*[\vec \eta]\e_a^*[\vec \xi]$, which is
\begin{equation}
    a(\vec\xi)a(\vec\eta)=a(S(R_{(2)}^{-1}(S(\xi),S(\eta))))a(S(R_{(1)}^{-1}(S(\xi),S(\eta))))
\end{equation}

Moving now to terms which contain both $a$ and $a^\dagger$, we focus on the expression
\begin{equation}
\label{eq:expr2}
    \int \frac{d^3\xi \, d^3\eta}{4\omega_{\vec \xi}\omega_{\vec \eta}}\left(a(\vec \xi)a^\dagger (\vec \eta)\e_a^*[\vec \xi]\e_b[\vec \eta]-a^\dagger(\vec \eta)a(\vec \xi)\e_b[\vec \eta]\e_a^*[\vec \xi]\right)
\end{equation}
Using techniques analogous to the ones that led us to \eqref{eq:adad}, we rewrite the second term of this integral as
\begin{equation}
\label{eq:aadint}\int_0^\infty d\xi_0\int d^3\xi \int_0^\infty d\eta_0 \int d^3\eta \, \, \delta(\xi^2-m^2)\delta(\eta^2-m^2) a^\dagger(\vec \eta)a(\vec \xi)\E_a[R_{(1)}(S(\xi),\eta)]\E_b[R_{(2)}(S(\xi),\eta)]
\end{equation}
We then perform the change of variables
\begin{equation}
    \label{eq:cov2}
    \begin{cases}
        S(\xi')=R_{(1)}(S(\xi),\eta)\\
        \eta'=R_{(2)}(S(\xi),\eta) 
    \end{cases}\rightarrow \quad  
    \begin{cases}
        \xi=S(R^{-1}_{(1)}(S(\xi'),\eta'))\\
        \eta=R^{-1}_{(2)}(S(\xi'),\eta') 
    \end{cases}
\end{equation}
where we have used the fact that $S[S(\xi)]=\xi$ for all unimodular models. The momentum space representations of the R-matrices are such that the change of variables \eqref{eq:cov2} satisfies the same properties outlined under the change of variables \eqref{eq:cov}. We can then rewrite \eqref{eq:aadint} as
\begin{equation}
\begin{aligned}
    &\int_0^\infty d\xi_0\int d^3\xi \int_0^\infty d\eta_0 \int d^3\eta \, \, \delta(\xi^2-m^2)\delta(\eta^2-m^2) a^\dagger(R^{-1}_{(2)}(S(\xi),\eta))a(S(R^{-1}_{(1)}(S(\xi),\eta)))\E_a[S(\xi)]\E_b[\eta]=\\
    =&\int \frac{d^3\xi d^3\eta}{4\omega_{\vec\xi}\omega_{\vec\eta}}a^\dagger(R^{-1}_{(2)}(S(\xi),\eta))a(S(R^{-1}_{(1)}(S(\xi),\eta)))\e_a^*[\vec\xi]\e_b[\vec\eta]
    \end{aligned}
\end{equation}
Thus, expression \eqref{eq:expr2} can be rewritten as
\begin{equation}
\label{eq:expr22}
    \int \frac{d^3\xi \, d^3\eta}{4\omega_{\vec \xi}\omega_{\vec \eta}}\left(a(\vec \xi)a^\dagger (\vec \eta)-a^\dagger(R^{-1}_{(2)}(S(\xi),\eta))a(S(R^{-1}_{(1)}(S(\xi),\eta))))\right)\e_a^*[\vec\xi]e_b[\vec\eta]
\end{equation}
and, comparing it to the Pauli-Jordan function, as written in the last equality of \eqref{eq:explicitexpansion}, yields the commutation relation
\begin{equation}
    a(\vec \xi)a^\dagger (\vec \eta)-a^\dagger(R^{-1}_{(2)}(S(\xi),\eta))a(S(R^{-1}_{(1)}(S(\xi),\eta))))=2\omega_{\vec\xi}\delta(\vec\xi-\vec\eta) \, .
\end{equation}
Notice that this last calculation also implicitly shows that
\begin{equation}
[\phi(\x_a),\phi(\x_b)]=W(\x_a-\x_b)
\end{equation}
which is \eqref{eq:phi+phi-} reported in the main text.


\begin{thebibliography}{10}

\bibitem{Vitale:2023znb}
P.~Vitale, M.~Adamo, R.~Dekhil, and D.~Fern{\'a}ndez-Silvestre \href{http://dx.doi.org/10.22323/1.440.0007}{{\em PoS} {\bfseries QG-MMSchools} (2024) 007}, \href{http://arxiv.org/abs/2309.17369}{{\ttfamily arXiv:2309.17369 [hep-th]}}.

\bibitem{Connes_book94}
A.~Connes, \href{http://dx.doi.org/10.4171/OWR/2007/43}{{\em Noncommutative geometry.}}
\newblock Oberwolfach Reports, 01, 1994.

\bibitem{MisnerWheeler}
C.~W. Misner and J.~A. Wheeler \href{http://dx.doi.org/https://doi.org/10.1016/0003-4916(57)90049-0}{{\em Ann. Phys.} {\bfseries 2} no.~6, (1957) 525--603}. \url{https://www.sciencedirect.com/science/article/pii/0003491657900490}.

\bibitem{Mattingly:2005re}
D.~Mattingly \href{http://dx.doi.org/10.12942/lrr-2005-5}{{\em Living Rev. Rel.} {\bfseries 8} (2005) 5}, \href{http://arxiv.org/abs/gr-qc/0502097}{{\ttfamily arXiv:gr-qc/0502097}}.

\bibitem{Bolmont:2022yad}
J.~Bolmont {\em et al.} \href{http://dx.doi.org/10.3847/1538-4357/ac5048}{{\em Astrophys. J.} {\bfseries 930} no.~1, (2022) 75}, \href{http://arxiv.org/abs/2201.02087}{{\ttfamily arXiv:2201.02087 [astro-ph.HE]}}.

\bibitem{AmelinoCamelia:2000mn}
G.~Amelino-Camelia \href{http://dx.doi.org/10.1142/S0218271802001330}{{\em Int. J. Mod. Phys.} {\bfseries D11} (2002) 35--60}, \href{http://arxiv.org/abs/gr-qc/0012051}{{\ttfamily arXiv:gr-qc/0012051 [gr-qc]}}.

\bibitem{majid_1995}
S.~Majid, \href{http://dx.doi.org/10.1017/CBO9780511613104}{{\em Foundations of Quantum Group Theory}}.
\newblock Cambridge University Press, 1995.

\bibitem{majid_2002}
S.~Majid, \href{http://dx.doi.org/10.1017/CBO9780511549892}{{\em A Quantum Groups Primer}}.
\newblock London Mathematical Society Lecture Note Series. Cambridge University Press, 2002.

\bibitem{Chari}
V.~Chari and A.~N. Pressley, {\em A guide to quantum groups}.
\newblock Cambridge University Press, 1994.

\bibitem{Mercati:2023apu}
F.~Mercati \href{http://dx.doi.org/10.1093/ptep/ptae088}{{\em PTEP} {\bfseries 2024} no.~7, (2024) 073B06}, \href{http://arxiv.org/abs/2311.16249}{{\ttfamily arXiv:2311.16249 [hep-th]}}.

\bibitem{Mercati:2024rzg}
F.~Mercati \href{http://dx.doi.org/10.1093/ptep/ptae175}{{\em PTEP} {\bfseries 2024} no.~12, (2024) 123B05}, \href{http://arxiv.org/abs/2404.08729}{{\ttfamily arXiv:2404.08729 [hep-th]}}.

\bibitem{Minwalla:1999px}
S.~Minwalla, M.~Van~Raamsdonk, and N.~Seiberg \href{http://dx.doi.org/10.1088/1126-6708/2000/02/020}{{\em JHEP} {\bfseries 02} (2000) 020}, \href{http://arxiv.org/abs/hep-th/9912072}{{\ttfamily arXiv:hep-th/9912072 [hep-th]}}.

\bibitem{Hersent:2023lqm}
K.~Hersent and J.-C. Wallet \href{http://dx.doi.org/10.1007/JHEP07(2023)031}{{\em JHEP} {\bfseries 07} (2023) 031}, \href{http://arxiv.org/abs/2304.05787}{{\ttfamily arXiv:2304.05787 [hep-th]}}.

\bibitem{Amelino-Camelia:2001rtw}
G.~Amelino-Camelia and M.~Arzano \href{http://dx.doi.org/10.1103/PhysRevD.65.084044}{{\em Phys. Rev. D} {\bfseries 65} (2002) 084044}, \href{http://arxiv.org/abs/hep-th/0105120}{{\ttfamily arXiv:hep-th/0105120}}.

\bibitem{Hersent:2024beg}
K.~Hersent, {\em {Field theories on quantum space-times : towards the phenomenology of quantum gravity}}.
\newblock PhD thesis, U. Paris-Saclay, 2024.
\newblock \href{http://arxiv.org/abs/2407.02023}{{\ttfamily arXiv:2407.02023 [math-ph]}}.

\bibitem{Grosse:2003nw}
H.~Grosse and R.~Wulkenhaar \href{http://dx.doi.org/10.1088/1126-6708/2003/12/019}{{\em JHEP} {\bfseries 12} (2003) 019}, \href{http://arxiv.org/abs/hep-th/0307017}{{\ttfamily arXiv:hep-th/0307017}}.

\bibitem{oeckl2000untwisting}
R.~Oeckl {\em Nuclear Physics B} {\bfseries 581} no.~1-2, (2000) 559--574.

\bibitem{Fiore:2007vg}
G.~Fiore and J.~Wess \href{http://dx.doi.org/10.1103/PhysRevD.75.105022}{{\em Phys. Rev. D} {\bfseries 75} (2007) 105022}, \href{http://arxiv.org/abs/hep-th/0701078}{{\ttfamily arXiv:hep-th/0701078}}.

\bibitem{wess1999qdeformed}
J.~Wess, ``q-Deformed Heisenberg Algebras,'' 1999.

\bibitem{Zakrzewski_1997}
S.~Zakrzewski {\em Communications in mathematical physics} {\bfseries 185} (1997) 285--311.

\bibitem{tolstoy2007twistedquantumdeformationslorentz}
V.~N. Tolstoy, ``Twisted Quantum Deformations of Lorentz and Poincar\'{e} algebras,'' 2007.
\newblock \url{https://arxiv.org/abs/0712.3962}.

\bibitem{Woronowicz:1989}
S.~Woronowicz \href{http://dx.doi.org/10.1007/BF0122141}{{\em Commun.Math. Phys.} {\bfseries 122} (1989) 125–170}.

\bibitem{Lizzi:2021rlb}
F.~Lizzi and F.~Mercati \href{http://dx.doi.org/10.1103/PhysRevD.103.126009}{{\em Phys. Rev. D} {\bfseries 103} (2021) 126009}, \href{http://arxiv.org/abs/2101.09683}{{\ttfamily arXiv:2101.09683 [hep-th]}}.

\bibitem{DiLuca:2022idu}
M.~G. Di~Luca and F.~Mercati \href{http://dx.doi.org/10.1103/PhysRevD.107.105018}{{\em Phys. Rev. D} {\bfseries 107} no.~10, (2023) 105018}, \href{http://arxiv.org/abs/2211.11627}{{\ttfamily arXiv:2211.11627 [hep-th]}}.

\bibitem{Agostini:2002de}
A.~Agostini, F.~Lizzi, and A.~Zampini \href{http://dx.doi.org/10.1142/S021773230200871X}{{\em Mod. Phys. Lett. A} {\bfseries 17} (2002) 2105--2126}, \href{http://arxiv.org/abs/hep-th/0209174}{{\ttfamily arXiv:hep-th/0209174}}.

\bibitem{Fabiano:2023xke}
G.~Fabiano and F.~Mercati \href{http://dx.doi.org/10.1103/PhysRevD.109.046011}{{\em Phys. Rev. D} {\bfseries 109} no.~4, (2024) 046011}, \href{http://arxiv.org/abs/2310.15063}{{\ttfamily arXiv:2310.15063 [hep-th]}}.

\bibitem{Drinfeld_thm}
V.~G. Drinfel'd {\em Leningrad Math. J.} {\bfseries 1} (1990) 321--342. \url{https://www.mathnet.ru/php/archive.phtml?wshow=paper&jrnid=aa&paperid=10&option_lang=eng}.

\bibitem{schweber2013introduction}
S.~Schweber, {\em An Introduction to Relativistic Quantum Field Theory}.
\newblock Dover Publications, 2013.

\bibitem{Bogdanovic:2024jnf}
D.~Bogdanovi{\'c}, M.~Dimitrijevi{\'c}~{\'C}iri{\'c}, V.~Radovanovi{\'c}, R.~J. Szabo, and G.~Trojani \href{http://dx.doi.org/10.1002/prop.202400169}{{\em Fortsch. Phys.} {\bfseries 72} no.~11, (2024) 2400169}, \href{http://arxiv.org/abs/2406.02372}{{\ttfamily arXiv:2406.02372 [hep-th]}}.

\bibitem{Lukierski:2005fc}
J.~Lukierski and M.~Woronowicz \href{http://dx.doi.org/10.1016/j.physletb.2005.11.052}{{\em Phys. Lett. B} {\bfseries 633} (2006) 116--124}, \href{http://arxiv.org/abs/hep-th/0508083}{{\ttfamily arXiv:hep-th/0508083}}.

\bibitem{Lizzi:2022hcq}
F.~Lizzi, L.~Scala, and P.~Vitale \href{http://dx.doi.org/10.1103/PhysRevD.106.025023}{{\em Phys. Rev. D} {\bfseries 106} no.~2, (2022) 025023}, \href{http://arxiv.org/abs/2205.10862}{{\ttfamily arXiv:2205.10862 [hep-th]}}.

\bibitem{Fabiano:2023uhg}
G.~Fabiano, G.~Gubitosi, F.~Lizzi, L.~Scala, and P.~Vitale \href{http://dx.doi.org/10.1007/JHEP08(2023)220}{{\em JHEP} {\bfseries 08} (2023) 220}, \href{http://arxiv.org/abs/2305.00526}{{\ttfamily arXiv:2305.00526 [hep-th]}}.

\bibitem{DimitrijevicCiric:2018blz}
M.~Dimitrijevic~Ciric, N.~Konjik, M.~A. Kurkov, F.~Lizzi, and P.~Vitale \href{http://dx.doi.org/10.1103/PhysRevD.98.085011}{{\em Phys. Rev. D} {\bfseries 98} no.~8, (2018) 085011}, \href{http://arxiv.org/abs/1806.06678}{{\ttfamily arXiv:1806.06678 [hep-th]}}.

\bibitem{Hayakawa:1999yt}
M.~Hayakawa \href{http://dx.doi.org/10.1016/S0370-2693(00)00242-2}{{\em Phys.\ Lett.\ B} {\bfseries 478} (2000) 394--400}, \href{http://arxiv.org/abs/hep-th/9912094}{{\ttfamily arXiv:hep-th/9912094 [hep-th]}}.

\bibitem{Gomis:2000zz}
J.~Gomis and T.~Mehen \href{http://dx.doi.org/10.1016/S0550-3213(00)00525-3}{{\em Nucl.\ Phys.\ B} {\bfseries 591} (2000) 265--276}, \href{http://arxiv.org/abs/hep-th/0005129}{{\ttfamily arXiv:hep-th/0005129 [hep-th]}}.

\bibitem{Matusis:2000jf}
A.~Matusis, L.~Susskind, and N.~Toumbas \href{http://dx.doi.org/10.1088/1126-6708/2000/12/002}{{\em JHEP} {\bfseries 12} (2000) 002}, \href{http://arxiv.org/abs/hep-th/0002075}{{\ttfamily arXiv:hep-th/0002075 [hep-th]}}.

\bibitem{Gubser:2000cd}
S.~S. Gubser and S.~L. Sondhi \href{http://dx.doi.org/10.1016/S0550-3213(01)00108-0}{{\em Nucl.\ Phys.\ B} {\bfseries 605} (2001) 395--424}, \href{http://arxiv.org/abs/hep-th/0006119}{{\ttfamily arXiv:hep-th/0006119 [hep-th]}}.

\bibitem{Carroll:2001ws}
S.~M. Carroll, J.~A. Harvey, V.~A. Kostelecký, C.~D. Lane, and T.~Okamoto \href{http://dx.doi.org/10.1103/PhysRevLett.87.141601}{{\em Phys.\ Rev.\ Lett.} {\bfseries 87} (2001) 141601}, \href{http://arxiv.org/abs/hep-th/0105082}{{\ttfamily arXiv:hep-th/0105082 [hep-th]}}.

\bibitem{Chaichian:2004yh}
M.~Chaichian, P.~Presnajder, and A.~Tureanu \href{http://dx.doi.org/10.1103/PhysRevLett.94.151602}{{\em Phys. Rev. Lett.} {\bfseries 94} (2005) 151602}, \href{http://arxiv.org/abs/hep-th/0409096}{{\ttfamily arXiv:hep-th/0409096}}.

\bibitem{Chaichian:2004za}
M.~Chaichian, P.~P. Kulish, K.~Nishijima, and A.~Tureanu \href{http://dx.doi.org/10.1016/j.physletb.2004.10.045}{{\em Phys. Lett. B} {\bfseries 604} (2004) 98--102}, \href{http://arxiv.org/abs/hep-th/0408069}{{\ttfamily arXiv:hep-th/0408069}}.

\bibitem{Giotopoulos:2021ieg}
G.~Giotopoulos and R.~J. Szabo \href{http://dx.doi.org/10.1088/1751-8121/ac5dad}{{\em J. Phys. A} {\bfseries 55} no.~35, (2022) 353001}, \href{http://arxiv.org/abs/2112.00541}{{\ttfamily arXiv:2112.00541 [hep-th]}}.

\bibitem{DimitrijevicCiric:2023hua}
M.~Dimitrijevi\'c~\'Ciri\'c, N.~Konjik, V.~Radovanovi\'c, and R.~J. Szabo \href{http://dx.doi.org/10.1007/JHEP08(2023)211}{{\em JHEP} {\bfseries 08} (2023) 211}, \href{http://arxiv.org/abs/2302.10713}{{\ttfamily arXiv:2302.10713 [hep-th]}}.

\bibitem{DimitrijevicCiric:2018}
M.~D. \'{C}iri\'{c}, N.~Konjik, M.~A. Kurkov, F.~Lizzi, and P.~Vitale \href{http://dx.doi.org/10.1103/PhysRevD.98.085011}{{\em Phys.\ Rev.\ D} {\bfseries 98} (2018) 085011}, \href{http://arxiv.org/abs/1806.06678}{{\ttfamily arXiv:1806.06678 [hep-th]}}.

\bibitem{HersentWallet:2023}
K.~Hersent and J.-C. Wallet \href{http://dx.doi.org/10.1007/JHEP07(2023)031}{{\em JHEP} {\bfseries 07} (2023) 031}, \href{http://arxiv.org/abs/2304.05787}{{\ttfamily arXiv:2304.05787 [hep-th]}}.

\bibitem{Hersent:2024}
K.~Hersent \href{http://dx.doi.org/10.1007/JHEP03(2024)023}{{\em JHEP} {\bfseries 03} (2024) 023}, \href{http://arxiv.org/abs/2309.08917}{{\ttfamily arXiv:2309.08917 [hep-th]}}.

\bibitem{MarisWallet:2024}
V.~Maris and J.-C. Wallet \href{http://dx.doi.org/10.1007/JHEP07(2024)119}{{\em JHEP} {\bfseries 07} (2024) 119}, \href{http://arxiv.org/abs/2401.18004}{{\ttfamily arXiv:2401.18004 [hep-th]}}.

\bibitem{Wallet:2025}
J.-C. Wallet \href{http://arxiv.org/abs/2503.07922}{{\ttfamily arXiv:2503.07922 [hep-th]}}. Preprint.

\bibitem{Oeckl:1999zu}
R.~Oeckl \href{http://dx.doi.org/10.1007/s002200100375}{{\em Commun. Math. Phys.} {\bfseries 217} (2001) 451--473}, \href{http://arxiv.org/abs/hep-th/9906225}{{\ttfamily arXiv:hep-th/9906225}}.

\bibitem{tolstoy2007twisted}
V.~N. Tolstoy in {\em Invited talk at the VII International Workshop `Lie Theory and its Applications in Physics',18--24 June 2007, Varna, Bulgaria}.
\newblock 2007.
\newblock \href{http://arxiv.org/abs/0712.3962}{{\ttfamily arXiv:0712.3962 [math.QA]}}.

\bibitem{Meier:2023kzt}
T.~Meier and S.~J. van Tongeren \href{http://dx.doi.org/10.1103/PhysRevLett.131.121603}{{\em Phys. Rev. Lett.} {\bfseries 131} no.~12, (2023) 121603}, \href{http://arxiv.org/abs/2301.08757}{{\ttfamily arXiv:2301.08757 [hep-th]}}.

\bibitem{Meier:2023lku}
T.~Meier and S.~J. van Tongeren \href{http://dx.doi.org/10.1007/JHEP12(2023)045}{{\em JHEP} {\bfseries 12} (2023) 045}, \href{http://arxiv.org/abs/2305.15470}{{\ttfamily arXiv:2305.15470 [hep-th]}}.

\end{thebibliography}
\end{document}